\documentclass[useAMS,usenatbib]{mn2e}

\usepackage{graphicx}
\graphicspath{{./fig/}}
\usepackage{bm}
\usepackage{url}
\title{Simulating field-aligned diffusion of a cosmic ray gas}

\author[Andrew P.\ Snodin et al.]%
{Andrew P.\ Snodin$^1$, Axel Brandenburg$^2$, Antony J.\ Mee$^1$, and Anvar Shukurov$^1$\\
$^1$School of Mathematics and Statistics, University of Newcastle,
  Newcastle upon Tyne, NE1 7RU, UK\\
$^2$NORDITA, Blegdamsvej 17, DK-2100 Copenhagen \O, Denmark}

\begin{document}


\newcommand{\EQ}{\begin{equation}}
\newcommand{\EN}{\end{equation}}
\newcommand{\EQA}{\begin{eqnarray}}
\newcommand{\ENA}{\end{eqnarray}}
\newcommand{\eq}[1]{(\ref{#1})}
\newcommand{\EEq}[1]{Equation~(\ref{#1})}
\newcommand{\Eq}[1]{Eq.~(\ref{#1})}
\newcommand{\Eqs}[2]{Eqs~(\ref{#1}) and~(\ref{#2})}
\newcommand{\Eqss}[2]{Eqs~(\ref{#1})--(\ref{#2})}
\newcommand{\eqs}[2]{(\ref{#1}) and~(\ref{#2})}
\newcommand{\App}[1]{Appendix~\ref{#1}}
\newcommand{\Sec}[1]{Sect.~\ref{#1}}
\newcommand{\Secs}[2]{Sects~\ref{#1} and \ref{#2}}
\newcommand{\Fig}[1]{Fig.~\ref{#1}}
\newcommand{\FFig}[1]{Figure~\ref{#1}}
\newcommand{\Tab}[1]{Table~\ref{#1}}
\newcommand{\Figs}[2]{Figs~\ref{#1} and \ref{#2}}
\newcommand{\Tabs}[2]{Tables~\ref{#1} and \ref{#2}}
\newcommand{\bra}[1]{\langle #1\rangle}
\newcommand{\bbra}[1]{\left\langle #1\right\rangle}
\newcommand{\mean}[1]{\overline{#1}}
\newcommand{\meanemf}{\overline{\mbox{\boldmath ${\cal E}$}}{}}{}
\newcommand{\meanemfs}{\overline{\cal E} {}}
\newcommand{\meanAA}{\overline{\bm{A}}}
\newcommand{\meanBB}{\overline{\bm{B}}}
\newcommand{\meanJJ}{\overline{\bm{J}}}
\newcommand{\meanUU}{\overline{\bm{U}}}
\newcommand{\meanWW}{\overline{\bm{W}}}
\newcommand{\meanFF}{\overline{{\cal{\bm{F}}}}}
\newcommand{\meanuu}{\overline{\mbox{\boldmath $u$}}{}}{}
\newcommand{\meanoo}{\overline{\mbox{\boldmath $\omega$}}{}}{}
\newcommand{\meanEE}{\overline{\mbox{\boldmath ${\cal E}$}}{}}{}
\newcommand{\meanuxB}{\overline{\mbox{\boldmath $\delta u\times \delta B$}}{}}{}
\newcommand{\meanJB}{\overline{\mbox{\boldmath $J\cdot B$}}{}}{}
\newcommand{\meanAB}{\overline{\mbox{\boldmath $A\cdot B$}}{}}{}
\newcommand{\meanjb}{\overline{\mbox{\boldmath $j\cdot b$}}{}}{}
\newcommand{\meanA}{\overline{A}}
\newcommand{\meanB}{\overline{B}}
\newcommand{\meanC}{\overline{C}}
\newcommand{\meanU}{\overline{U}}
\newcommand{\meanJ}{\overline{J}}
\newcommand{\meanS}{\overline{S}}
\newcommand{\meanF}{\overline{\cal F}}
%
%
\newcommand{\teps}{\tilde{\epsilon} {}}
\newcommand{\zh}{\widehat{z}}
%
%
\newcommand{\eee}{\widehat{\mbox{\boldmath $e$}} {}}
\newcommand{\nnn}{\widehat{\mbox{\boldmath $n$}} {}}
\newcommand{\vvv}{\widehat{\mbox{\boldmath $v$}} {}}
\newcommand{\rr}{\widehat{\mbox{\boldmath $r$}} {}}
\newcommand{\xxx}{\widehat{\bm x}}
\newcommand{\yyy}{\widehat{\bm y}}
\newcommand{\zz}{\widehat{\bm z}}
\newcommand{\pphi}{\widehat{\bm\phi}}
\newcommand{\ttt}{\widehat{\mbox{\boldmath $\theta$}} {}}
\newcommand{\OOO}{\widehat{\mbox{\boldmath $\Omega$}} {}}
\newcommand{\ooo}{\widehat{\mbox{\boldmath $\omega$}} {}}
\newcommand{\BBBB}{\widehat{\mbox{\boldmath $B$}} {}}
\newcommand{\Bhat}{\widehat{B}}
\newcommand{\BBhat}{\widehat{\bm{B}}}
%
%
\newcommand{\gggg}{\mbox{\boldmath $g$} {}}
\newcommand{\ddd}{\mbox{\boldmath $d$} {}}
\newcommand{\rrr}{\mbox{\boldmath $r$} {}}
\newcommand{\yy}{\mbox{\boldmath $y$} {}}
\newcommand{\zzz}{\mbox{\boldmath $z$} {}}
\newcommand{\vv}{\mbox{\boldmath $v$} {}}
\newcommand{\ww}{\mbox{\boldmath $w$} {}}
\newcommand{\mm}{\mbox{\boldmath $m$} {}}
\newcommand{\PP}{\mbox{\boldmath $P$} {}}
\newcommand{\bp}{\mbox{\boldmath $p$} {}}
\newcommand{\pp}{\mbox{\boldmath $p$} {}}
\newcommand{\II}{\mbox{\boldmath $I$} {}}
\newcommand{\qq}{{\bm{q}}}
\newcommand{\xx}{{\bm{x}}}
\newcommand{\UU}{{\bm{U}}}
\newcommand{\WW}{{\bm{W}}}
\newcommand{\QQ}{{\bm{Q}}}
\newcommand{\uu}{{\bm{u}}}
\newcommand{\BB}{{\bm{B}}}
\newcommand{\HH}{{\bm{H}}}
\newcommand{\CC}{{\bm{C}}}
\newcommand{\JJ}{{\bm{J}}}
\newcommand{\jj}{{\bm{j}}}
\newcommand{\AAA}{{\bm{A}}}
\newcommand{\aaaa}{{\bm{a}}}
\newcommand{\bb}{{\bm{b}}}
\newcommand{\cc}{{\bm{c}}}
\newcommand{\ee}{{\bm{e}}}
\newcommand{\nn}{\mbox{\boldmath $n$} {}}
\newcommand{\ff}{\mbox{\boldmath $f$} {}}
\newcommand{\hh}{\mbox{\boldmath $h$} {}}
\newcommand{\EE}{{\bm{E}}}
\newcommand{\FF}{{\bm{F}}}
\newcommand{\FFF}{{\bm{{\cal F}}}}
\newcommand{\KK}{{\bm{K}}}
\newcommand{\kk}{{\bm{k}}}
\newcommand{\TT}{\mbox{\boldmath $T$} {}}
\newcommand{\MM}{\mbox{\boldmath $M$} {}}
\newcommand{\GG}{\mbox{\boldmath $G$} {}}
\newcommand{\SSS}{{\bm{S}}}
\newcommand{\grav}{\mbox{\boldmath $g$} {}}
\newcommand{\nab}{\mbox{\boldmath $\nabla$} {}}
\newcommand{\OO}{\mbox{\boldmath $\Omega$} {}}
\newcommand{\oo}{\mbox{\boldmath $\omega$} {}}
\newcommand{\ttau}{\mbox{\boldmath $\tau$} {}}
\newcommand{\LL}{\mbox{\boldmath $\Lambda$} {}}
\newcommand{\llambda}{\mbox{\boldmath $\lambda$} {}}
\newcommand{\pomega}{\mbox{\boldmath $\varpi$} {}}
%
%
\newcommand{\SSSS}{\mbox{\boldmath $\sf S$} {}}
\newcommand{\RRRR}{\mbox{\boldmath $\sf R$} {}}
\newcommand{\LLLL}{\mbox{\boldmath $\sf L$} {}}
\newcommand{\PPPP}{\mbox{\boldmath ${\sf P}$} {}}
\newcommand{\MMMM}{\mbox{\boldmath ${\sf M}$} {}}
\newcommand{\AAAA}{\mbox{\boldmath ${\cal A}$} {}}
\newcommand{\BBB}{\mbox{\boldmath ${\cal B}$} {}}
\newcommand{\emf}{\mbox{\boldmath ${\cal E}$} {}}
\newcommand{\GGG}{\mbox{\boldmath ${\cal G}$} {}}
\newcommand{\HHH}{\mbox{\boldmath ${\cal H}$} {}}
\newcommand{\QQQ}{\mbox{\boldmath ${\cal Q}$} {}}
\newcommand{\GGGG}{{\bf G} {}}
%
%
\newcommand{\ii}{{\rm i}}
\newcommand{\grad}{{\rm grad} \, {}}
\newcommand{\curl}{{\rm curl} \, {}}
\newcommand{\dive}{{\rm div}  \, {}}
\newcommand{\Dive}{{\rm Div}  \, {}}
\newcommand{\sgn}{{\rm sgn}  \, {}}
\newcommand{\DD}{{\rm D} {}}
\newcommand{\DDD}{{\cal D} {}}
\newcommand{\dd}{{\rm d} {}}
\newcommand{\const}{{\rm const}  {}}
\newcommand{\CR}{{\rm CR}}
\def\degr{\hbox{$^\circ$}}
\def\la{\mathrel{\mathchoice {\vcenter{\offinterlineskip\halign{\hfil
$\displaystyle##$\hfil\cr<\cr\sim\cr}}}
{\vcenter{\offinterlineskip\halign{\hfil$\textstyle##$\hfil\cr<\cr\sim\cr}}}
{\vcenter{\offinterlineskip\halign{\hfil$\scriptstyle##$\hfil\cr<\cr\sim\cr}}}
{\vcenter{\offinterlineskip\halign{\hfil$\scriptscriptstyle##$\hfil\cr<\cr\sim\cr}}}}}
\def\ga{\mathrel{\mathchoice {\vcenter{\offinterlineskip\halign{\hfil
$\displaystyle##$\hfil\cr>\cr\sim\cr}}}
{\vcenter{\offinterlineskip\halign{\hfil$\textstyle##$\hfil\cr>\cr\sim\cr}}}
{\vcenter{\offinterlineskip\halign{\hfil$\scriptstyle##$\hfil\cr>\cr\sim\cr}}}
{\vcenter{\offinterlineskip\halign{\hfil$\scriptscriptstyle##$\hfil\cr>\cr\sim\cr}}}}}
%
%
\def\Ta{\mbox{\rm Ta}}
\def\Ra{\mbox{\rm Ra}}
\def\Ma{\mbox{\rm Ma}}
\def\Co{\mbox{\rm Co}}
\def\Roo{\mbox{\rm Ro}^{-1}}
\def\Rooo{\mbox{\rm Ro}^{-2}}
\def\Pra{\mbox{\rm Pr}}
\def\Pran{\mbox{\rm Pr}}
\def\Pm{\mbox{\rm Pr}_M}
\def\Rm{R_\mathrm{m}}
\def\Rey{\mbox{\rm Re}}
\def\Pe{\mbox{\rm Pe}}
\newcommand{\ea}{{\rm et al.\ }}
\newcommand{\eaa}{{\rm et al.\ }}
\def\half{{\textstyle{1\over2}}}
\def\threehalf{{\textstyle{3\over2}}}
\def\onethird{{\textstyle{1\over3}}}
\def\onesixth{{\textstyle{1\over6}}}
\def\twothird{{\textstyle{2\over3}}}
\def\fourthird{{\textstyle{4\over3}}}
\def\quarter{{\textstyle{1\over4}}}
\newcommand{\W}{\,{\rm W}}
\newcommand{\V}{\,{\rm V}}
\newcommand{\kV}{\,{\rm kV}}
\newcommand{\T}{\,{\rm T}}
\newcommand{\G}{\,{\rm G}}
\newcommand{\Hz}{\,{\rm Hz}}
\newcommand{\nHz}{\,{\rm nHz}}
\newcommand{\kHz}{\,{\rm kHz}}
\newcommand{\kG}{\,{\rm kG}}
\newcommand{\mkG}{\,\mu{\rm G}}
\newcommand{\K}{\,{\rm K}}
\newcommand{\g}{\,{\rm g}}
\newcommand{\s}{\,{\rm s}}
\newcommand{\ms}{\,{\rm ms}}
\newcommand{\mpers}{\,{\rm m/s}}
\newcommand{\ks}{\,{\rm ks}}
\newcommand{\cm}{\,{\rm cm}}
\newcommand{\m}{\,{\rm m}}
\newcommand{\km}{\,{\rm km}}
\newcommand{\kms}{\,{\rm km/s}}
\newcommand{\kg}{\,{\rm kg}}
\newcommand{\ug}{\,\mu{\rm g}}
\newcommand{\kW}{\,{\rm kW}}
\newcommand{\MW}{\,{\rm MW}}
\newcommand{\Mm}{\,{\rm Mm}}
\newcommand{\Mx}{\,{\rm Mx}}
\newcommand{\p}{\,{\rm pc}}
\newcommand{\kpc}{\,{\rm kpc}}
\newcommand{\yr}{\,{\rm yr}}
\newcommand{\Myr}{\,{\rm Myr}}
\newcommand{\Gyr}{\,{\rm Gyr}}
\newcommand{\erg}{\,{\rm erg}}
\newcommand{\mol}{\,{\rm mol}}
\newcommand{\dyn}{\,{\rm dyn}}
\newcommand{\J}{\,{\rm J}}
\newcommand{\RM}{\,{\rm RM}}
\newcommand{\EM}{\,{\rm EM}}
\newcommand{\AU}{\,{\rm AU}}
\newcommand{\A}{\,{\rm A}}
%
%
\newcommand{\yastroph}[2]{ #1, astro-ph/#2}
\newcommand{\ycsf}[3]{ #1, {Chaos, Solitons \& Fractals,} {#2}, #3}
\newcommand{\yepl}[3]{ #1, {Europhys. Lett.,} {#2}, #3}
\newcommand{\yaj}[3]{ #1, {AJ,} {#2}, #3}
\newcommand{\yjgr}[3]{ #1, {JGR,} {#2}, #3}
\newcommand{\ysol}[3]{ #1, {Sol. Phys.,} {#2}, #3}
\newcommand{\yapj}[3]{ #1, {ApJ,} {#2}, #3}
\newcommand{\ypasp}[3]{ #1, {PASP,} {#2}, #3}
\newcommand{\yapjl}[3]{ #1, {ApJ,} {#2}, #3}
\newcommand{\yapjs}[3]{ #1, {ApJS,} {#2}, #3}
\newcommand{\yan}[3]{ #1, {AN,} {#2}, #3}
\newcommand{\yzfa}[3]{ #1, {Z.\ f.\ Ap.,} {#2}, #3}
\newcommand{\ymhdn}[3]{ #1, {Magnetohydrodyn.} {#2}, #3}
\newcommand{\yana}[3]{ #1, {A\&A,} {#2}, #3}
\newcommand{\yanas}[3]{ #1, {A\&AS,} {#2}, #3}
\newcommand{\yanar}[3]{ #1, {A\&AR,} {#2}, #3}
\newcommand{\yass}[3]{ #1, {Ap\&SS,} {#2}, #3}
\newcommand{\ygafd}[3]{ #1, {Geophys. Astrophys. Fluid Dyn.,} {#2}, #3}
\newcommand{\ypasj}[3]{ #1, {Publ. Astron. Soc. Japan,} {#2}, #3}
\newcommand{\yjfm}[3]{ #1, {JFM,} {#2}, #3}
\newcommand{\ypf}[3]{ #1, {Phys. Fluids,} {#2}, #3}
\newcommand{\ypp}[3]{ #1, {Phys. Plasmas,} {#2}, #3}
\newcommand{\ysov}[3]{ #1, {Sov. Astron.,} {#2}, #3}
\newcommand{\ysovl}[3]{ #1, {Sov. Astron. Lett.,} {#2}, #3}
\newcommand{\yjetp}[3]{ #1, {Sov. Phys. JETP,} {#2}, #3}
\newcommand{\yphy}[3]{ #1, {Physica,} {#2}, #3}
\newcommand{\yannr}[3]{ #1, {ARA\&A,} {#2}, #3}
\newcommand{\yaraa}[3]{ #1, {ARA\&A,} {#2}, #3}
\newcommand{\yprs}[3]{ #1, {Proc. Roy. Soc. Lond.,} {#2}, #3}
\newcommand{\yprl}[3]{ #1, {Phys. Rev. Lett.,} {#2}, #3}
\newcommand{\yphl}[3]{ #1, {Phys. Lett.,} {#2}, #3}
\newcommand{\yptrs}[3]{ #1, {Phil. Trans. Roy. Soc.,} {#2}, #3}
\newcommand{\ymn}[3]{ #1, {MNRAS,} {#2}, #3}
\newcommand{\ynat}[3]{ #1, {Nat,} {#2}, #3}
\newcommand{\ysci}[3]{ #1, {Sci,} {#2}, #3}
\newcommand{\ysph}[3]{ #1, {Solar Phys.,} {#2}, #3}
\newcommand{\ypr}[3]{ #1, {Phys. Rev.,} {#2}, #3}
\newcommand{\ypre}[3]{ #1, {Phys. Rev. E,} {#2}, #3}
\newcommand{\spr}[2]{ ~#1~ {\em Phys. Rev. }{\bf #2} (submitted)}
\newcommand{\ppr}[2]{ ~#1~ {\em Phys. Rev. }{\bf #2} (in press)}
\newcommand{\ypnas}[3]{ #1, {Proc. Nat. Acad. Sci.,} {#2}, #3}
\newcommand{\yicarus}[3]{ #1, {Icarus,} {#2}, #3}
\newcommand{\yspd}[3]{ #1, {Sov. Phys. Dokl.,} {#2}, #3}
\newcommand{\yjcp}[3]{ #1, {J. Comput. Phys.,} {#2}, #3}
\newcommand{\yjour}[4]{ #1, {#2}, {#3}, #4}
\newcommand{\yprep}[2]{ #1, {\sf #2}}
\newcommand{\ybook}[3]{ #1, {#2} (#3)}
\newcommand{\yproc}[5]{ #1, in {#3}, ed. #4 (#5), #2}
\newcommand{\pproc}[4]{ #1, in {#2}, ed. #3 (#4), (in press)}
\newcommand{\ppp}[1]{ #1, {Phys. Plasmas,} (in press)}
\newcommand{\sapj}[1]{ #1, {ApJ,} (submitted)}
\newcommand{\sana}[1]{ #1, {A\&A,} (submitted)}
\newcommand{\san}[1]{ #1, {AN,} (submitted)}
\newcommand{\sprl}[1]{ #1, {PRL,} (submitted)}
\newcommand{\pprl}[1]{ #1, {PRL,} (in press)}
\newcommand{\sjfm}[1]{ #1, {JFM,} (submitted)}
\newcommand{\sgafd}[1]{ #1, {Geophys. Astrophys. Fluid Dyn.,} (submitted)}
\newcommand{\pgafd}[1]{ #1, {Geophys. Astrophys. Fluid Dyn.,} (in press)}
\newcommand{\tana}[1]{ #1, {A\&A,} (to be submitted)}
\newcommand{\smn}[1]{ #1, {MNRAS,} (submitted)}
\newcommand{\pmn}[1]{ #1, {MNRAS,} (in press)}
\newcommand{\papj}[1]{ #1, {ApJ,} (in press)}
\newcommand{\papjl}[1]{ #1, {ApJL,} (in press)}
\newcommand{\sapjl}[1]{ #1, {ApJL,} (submitted)}
\newcommand{\pana}[1]{ #1, {A\&A,} (in press)}
\newcommand{\pan}[1]{ #1, {AN,} (in press)}
\newcommand{\pjour}[2]{ #1, {#2,} (in press)}
\newcommand{\sjour}[2]{ #1, {#2,} (submitted)}

\date{\today,~ $ $Revision: 1.242 $ $}
\pagerange{\pageref{firstpage}--\pageref{lastpage}}
\pubyear{2006}
\maketitle
\label{firstpage}

\begin{abstract}
The macroscopic behaviour of
cosmic rays in turbulent magnetic fields is discussed.
An implementation of anisotropic diffusion of cosmic rays with respect to
the magnetic field in a non-conservative, high-order, finite-difference
magnetohydrodynamic code is discussed.
It is shown that the standard implementation fails
near singular {\sf X}-points of the magnetic field, which are
common if the field is random.
A modification to the diffusion model for cosmic rays is described
and the resulting telegraph equation (implemented by
solving a dynamic equation for the diffusive flux of cosmic rays)
is used; it is argued that this modification may better describe the physics of cosmic ray diffusion.
The present model reproduces several processes important for the propagation
and local confinement of
cosmic rays, including spreading perpendicular to the local large-scale
magnetic field, controlled by the random-to-total magnetic field ratio,
and the balance between cosmic ray pressure and magnetic tension.
Cosmic ray diffusion is discussed in the context of a random
magnetic field produced by turbulent dynamo action.
It is argued that
energy equipartition between cosmic rays and other constituents
of the interstellar medium do not necessarily imply that cosmic rays play
a significant role in the balance of forces.
\end{abstract}

\section{Introduction}

The importance of cosmic rays for the dynamics of the interstellar medium (ISM)
has long been recognized \citep{Parker1966,Berezinskiietal90}.
Spatial gradients of the cosmic ray pressure contribute significantly
to the force balance in the ISM.
If cosmic rays are confined within magnetic flux tubes, then the tendency
toward pressure equilibrium reduces gas pressure within the tubes.
Depending on the efficiency of cooling,
either temperature or entropy will be approximately uniform
across the tube, but in both cases density inside the tube will be
decreased relative to the exterior, making the tube buoyant.
This process is similar to magnetic buoyancy.
Therefore, cosmic rays
facilitate disc--halo connections in spiral galaxies by enhancing the buoyancy
of magnetic structures in the interstellar gas.
In the sun, magnetic buoyancy drives magnetic flux tubes to the
surface to form bipolar regions.
In galaxies, magnetic buoyancy is believed to be strongly assisted
by cosmic rays.

The effects of cosmic-ray driven buoyancy are believed to be important
for the operation of the galactic dynamo \citep{Parker1992,MSS99}.
This can help to speed up the growth of the magnetic field and
maintain strong field amplification and regeneration, especially in the
non-linear regime \citep{HanaszEtal2004}.
Many studies of the Parker instability as well as recent simulations of
the galactic dynamo rely on a hydrodynamic description of cosmic rays
\citep{SL85},
which is especially convenient in models involving the large-scale dynamics
of the interstellar medium.
In this approach, the cosmic ray transport equation for the phase space
distribution function is integrated over particle momenta
which results in a hydrodynamic-type equation for the cosmic ray energy
density or pressure.
Our aim here is to use this approach in order to clarify the relation between
cosmic ray energy density and properties of the interstellar medium.

Energy equipartition (or pressure balance) between cosmic rays and magnetic fields
is a common assumption used in radio astronomy, where it is
used to estimate magnetic field strength from synchrotron intensity.
A physical basis for this idea remains elusive and only
qualitative arguments, related to cosmic ray confinement by magnetic fields, are
used to justify this concept. The assumption comes into question since the spatial
distribution of cosmic rays may not precisely follow that of magnetic field strength.
Furthermore, the idea of
overall (statistical) pressure balance in the ISM would be more difficult to
maintain if both magnetic
and cosmic ray pressures are enhanced or reduced at the same positions
simultaneously.
Recent arguments of \cite{PadoanScalo05} suggest that, if the
streaming velocity of cosmic rays is proportional to the Alfv\'en speed
\citep[][and references therein]{FK01,FG04}, the local cosmic ray density
is independent of the local magnetic field strength,
and rather scales with the square root of the (ionized) gas density.
Indeed, if both the magnetic flux and the cosmic ray flux are
conserved, $BS=\mathrm{const}$ and $n_\mathrm{c}US=\mathrm{const}$ (where
$B$ is the magnetic field strength, $S$ is the area within a fluid contour,
$n_\mathrm{c}$ is the number density of cosmic rays and $U$ is their streaming
velocity), one obtains $n_\mathrm{c}U/B=\mathrm{const}$, which yields
$n_\mathrm{c}\propto n_\mathrm{i}^{1/2}$, given that $U= V_\mathrm{A}\propto B n_\mathrm{i}^{-1/2}$,
with $n_\mathrm{i}$ the ion number density and $V_\mathrm{A}$ the Alfv\'en speed.

We use a two-fluid model, where
cosmic rays are described by an equation for their pressure (or
energy density) and an equation of state.
The cosmic rays are assumed to act directly
on the background gas via their pressure gradient.
We do not include any explicit means of exciting hydromagnetic
waves by cosmic rays leading to their confinement
\citep[for a discussion of confinement issues, see][]{C80},
but instead parameterize these processes by choosing an appropriate advection velocity
(as a superposition of the gas and Alfv\'en velocities).
There are several interesting questions regarding
high energy cosmic rays and their acceleration (e.g.\ Hillas 2005),
which we are not attempting to address here.
Instead, we want to know which process is mainly responsible for
limiting the cosmic rays energy density and what is the
relation of cosmic ray energy density with the magnetic field.
Is there local equipartition, or is there only global equipartition
on the scale of the galaxy?
Finally, we are interested in studying those effects in the ISM dynamics
that only arise in the presence of cosmic rays.
We begin with the governing equations and discuss issues that arise
in connection with the numerical implementation of cosmic ray diffusion
along magnetic field lines.

\section{Method}
\subsection{Basic equations}

The hydromagnetic equations, supplemented by the advection-diffusion
equation for the cosmic ray energy density, and the cosmic ray pressure
in the momentum equation, are
\EQ
{\partial\rho\over\partial t}+\nab\cdot(\rho\uu)=0,
\EN
\EQ
{\partial e_{\rm c}\over\partial t}+\nab\cdot(e_{\rm c}\uu)
+p_{\rm c}\nab\cdot\uu=D_{\rm c}+Q_{\rm c},
\label{decdt}
\EN
\EQ
{\partial e_{\rm g}\over\partial t}+\nab\cdot(e_{\rm g}\uu)
+p_{\rm g}\nab\cdot\uu=D_{\rm g}+Q_{\rm k}+Q_{\rm m},
\EN
\EQ \label{drudt}
{\partial\rho\uu\over\partial t}+\nab\cdot(\rho\uu\uu)
+\nab(p_{\rm g}+p_{\rm c})=\JJ\times\BB+\ff+\FF,
\EN
\EQ
{\partial\BB\over\partial t}=\nab\times(\uu\times\BB-\eta\mu_0\JJ),
\EN
where $\rho$, $\uu$ and $p_\mathrm{g}$ are the gas density, velocity and pressure;
$e_\mathrm{c}$ and $p_\mathrm{c}$ are the cosmic ray energy density and pressure,
$\BB$ is the magnetic field,
$\JJ=\nab\times\BB/\mu_0$ is the electric current density,
$\eta$ is the magnetic diffusivity,
$D_{\rm g}=\nab\cdot(K\nab T)$ is the thermal diffusion term
(treated here isotropically; thermal diffusion is unimportant in
the present context, but weak diffusion is necessary for numerical reasons).
Further, $T$ is the temperature
related to the internal energy density (per unit volume),
$e_{\rm g}$, via $e_{\rm g}=\rho c_v T$,
and $D_{\rm c}$ is the divergence of the diffusive cosmic
ray energy flux taken with the opposite sign, i.e.\
\EQ
D_{\rm c}=-\nab\cdot\FFF_{\rm c}.
\label{FluxDivergence}
\EN
The usual approach is to treat this term as Fickian diffusion, i.e.,
to assume that the flux is proportional to the instantaneous gradient
of the cosmic ray energy density,
\EQ
{\cal F}_{{\rm c}i}=-K_{ij}\nabla_j e_{\rm c}
\quad\mbox{(Fickian diffusion)},
\label{Fickian}
\EN
where $K_{ij}$ is the diffusion tensor.
The latter can be written as
\EQ
K_{ij}=K_\perp\delta_{ij}+(K_\parallel-K_\perp)\Bhat_i\Bhat_j,
\label{Difftensor}
\EN
where $\BBhat=\BB/|\BB|$ is the field-aligned unit vector
\citep[e.g.][]{Berezinskiietal90,HanaszLesch2003}.
Here, $K_\parallel$ and $K_\perp$ are the cosmic ray diffusion
coefficients along and perpendicular to the field, respectively.

We assume ideal-gas equations of state for both the cosmic rays and
the gas, i.e.\ $p_{\rm c}=(\gamma_{\rm c}-1)e_{\rm c}$ and
$p_{\rm g}=(\gamma_{\rm g}-1)e_{\rm g}$, where $\gamma_{\rm c}$
and $\gamma_{\rm g}$ are the ratios of the total number of degrees of
freedom to the number of translational degrees of freedom for the
cosmic rays and the gas.
Unless stated otherwise, we assume $\gamma_{\rm c}=4/3$ and $\gamma_{\rm g}=5/3$.
Other choices for $\gamma_{\rm c}$ include $5/3$ and $14/9$
\citep[e.g.][and references therein]{RYU}.

The system can be driven by an external force $\ff$ in the momentum equation (\ref{drudt}),
and $\FF$ in that equation includes additional forces such as
the viscous force, $\nab\cdot(2\nu\rho\SSSS)$, where
$\nu$ is the viscosity and ${\sf S}_{ij}=\half(u_{i,j}+u_{j,i})
-\onethird\delta_{ij}u_{k,k}$ is the traceless rate of strain tensor,
where commas denote partial differentiation.
Furthermore, $Q_{\rm k}={2\rho\nu\SSSS^2}$ and
$Q_{\rm m}={\eta\mu_0\JJ^2}$ denote the viscous and Joule
heating, and $Q_{\rm c}$ is a
cosmic ray energy source.

\subsection{Non-Fickian diffusion}
\label{NonFickian}

Typical values of the diffusivity along the magnetic field are of the order $10^{28}\cm^2\s^{-1}$
\citep[e.g.][]{Berezinskiietal90}.
Such large values would severely limit numerical modelling since a large
diffusivity requires that the computational time step is small to ensure numerical
stability; for example simulations with a resolution of 1 pc would require a
time step of 10 years or less
(e.g., Hanasz \& Lesch 2003 reduce $K_\parallel$ by a factor of 10
to make the system tractable numerically).
This problem could be circumvented by employing an implicit numerical scheme.
In the context of cosmic ray propagation, one would expect
the advection speed to be not too much larger than the
Alfv\'en speed.
Before discussing a possible remedy to this problem we note that, in
the case of field-aligned diffusion, the problem can be even more severe.
If we use the product rule and
write $D_{\rm c}=\nabla_i(K_{ij}\nabla_j e_{\rm c})$ in the form
\EQ
D_{\rm c}=-\UU_{\rm c}\cdot\nab e_{\rm c}+K_{ij}\partial_i\partial_j e_{\rm c},
\label{Dc}
\EN
we see that
$U_{{\rm c}\,i}=-\partial K_{ij}/\partial x_j$
plays the role of a
velocity transporting cosmic rays perpendicular to curved
field lines.
This term is proportional to the divergence of the
dyadic product of unit vectors, $\nab\cdot(\BBhat\BBhat)$.
At magnetic {\sf X}-points,
this term is singular, as explained below
(we note that {\sf O}-type singular magnetic points do not cause difficulties).

We illustrate this complication using
a simple magnetic field configuration  $\BB=(x,-y,0)^T$ with a null point at the
origin, which leads to the singular behaviour
of $\nab\cdot(\BBhat\BBhat)$, and hence to a singularity of $|\UU_{\rm c}|$:
\EQ
\nab\cdot(\BBhat\BBhat)={1\over r^4}\pmatrix{(3y^2-x^2)x\cr(3x^2-y^2)y\cr0},
\EN
where $r^2=x^2+y^2$.
This expression diverges
at the origin and leads to infinite propagation speed
which would, technically speaking, limit to zero the length of the time step
of an explicit timestepping scheme.
In spite of this singularity, the cosmic ray energy density must
stay finite.
In fact, one can show that, in a closed or periodic domain,
the maximum
cosmic ray energy density, $\max(e_{\rm c})$,
can only decrease with time.  This is a
well-known general property of the diffusion operator; in \App{Bounded}
we derive this result for the form of the diffusion tensor
appropriate for cosmic rays.
The reason that $\max(e_{\rm c})$ can remain finite, despite
$\nab\cdot(\BBhat\BBhat)$, and hence $\UU_{\rm c}$, becoming infinite,
is that the parabolic system of equations can adjust itself instantaneously
so that $\nab e_{\rm c}$ tends to zero where $\UU_{\rm c}$ diverges.
A numerically convenient remedy to this problem will be discussed in \Sec{Xpoint},
where a non-Fickian diffusion model is used.
In the following we describe this approach in more detail.

A physically appealing, and widely adopted way to improve the diffusion
equation so as to limit the propagation speed to a finite value involves a more
accurate description of the diffusive flux. This generalization has been applied,
e.g., to turbulent diffusion.
In turbulence, the classical turbulent diffusion
equation, $\partial n/\partial t=D\,\partial^2 n/\partial x^2$, arises if the
turbulent velocity field is assumed to be $\delta$-correlated in time; this
approximation is consistent with Eq.~(\ref{Fickian}) or its simplifications.
In order to ensure finite propagation speed of the diffusing substance, it is
sufficient to allow for a finite correlation time $\tau$ of the velocity field.
This leads to equation (\ref{nonFickian}) for the diffusive flux. The
corresponding equation for the diffusing quantity reduces to
the telegraph equation
$\partial n/\partial t+\tau\,\partial^2 n/\partial t^2=D\,\partial^2 n/\partial x^2$,
or its generalizations. These arguments have been recently discussed by
\citet{B03a,B03b}.
The telegraph equation has been used to
correct acausal cosmic-ray diffusion models \citep[e.g.,][]{GombosiEtal93}.
This type of non-Fickian diffusion also emerges quite naturally in
turbulent diffusion of passive scalars \citep{BF03} and has been confirmed
in direct simulations \citep{BKM04}.
On long enough time scales, or for sufficiently small values of
$\tau$, the non-Fickian description of diffusion reduces to the Fickian limit.

Thus, we replace \Eq{Fickian} by
\EQ
{\partial{\cal F}_{{\rm c}i}\over\partial t}=-\tilde{K}_{ij}\nabla_je_{\rm c}
-{{\cal F}_{{\rm c}i}\over\tau}
\quad\mbox{(non-Fickian diffusion)},
\label{nonFickian}
\EN
where $K_{ij}=\tau\tilde{K}_{ij}$ corresponds to the original diffusion tensor.
Similarly to \Eq{Difftensor}, we write
\EQ
\tilde{K}_{ij}=\tilde{K}_\perp\delta_{ij}
+(\tilde{K}_\parallel-\tilde{K}_\perp)\Bhat_i\Bhat_j.
\EN
Quantitatively, the deviation from Fick's law is controlled by the dimensionless parameter
\EQ
\mbox{St}=\frac{\tilde{K}_\parallel^{1/2}\tau}{\ell}
=\frac{(K_\parallel\tau)^{1/2}}{\ell},
\label{Stdef}
\EN
where $\ell$ is the typical length scale of the initial structure.
In the context of turbulent diffusion, this dimensionless parameter
is often referred to as the Strouhal number \citep{LL87,KR80}.
The larger the
Strouhal number, the more important are non-Fickian effects
resulting in a wave-like behaviour of the solution.
Unlike the solution of the classical diffusion equation, where an initial
perturbation to the trivial solution
has an effect at every position for any $t>0$, solutions with non-Fickian
diffusion remain unperturbed ahead of a propagating front.

\begin{figure}\begin{center}
\includegraphics[width=.4\textwidth]{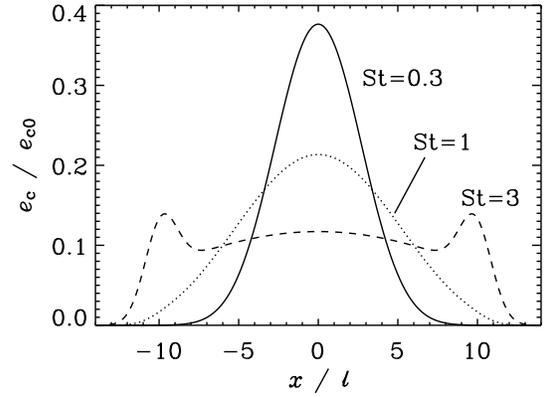}
\end{center}\caption[]{
The spread of an initial Gaussian distribution of cosmic ray energy density
(of a half-width $\ell$): the distribution at a time $t=1$ is shown,
as a function of $x/\ell$,
for three values of the Strouhal number St. Note that the behaviour of the solution
becomes more wave-like as St increases.
}\label{ptau}\end{figure}

A suitable estimate of the Strouhal number can be obtained assuming that the
relevant correlation time is of the order of the Alfv\'en crossing time for magnetic
structures of scale $\ell$, i.e.\
$\mathrm{St}\simeq(K_\parallel/V_\mathrm{A}\ell)^{1/2}$.
This yields (for gas number density $0.1\cm^{-3}$)
\begin{equation}\label{Stest}
\mathrm{St}\simeq20\left(\frac{K_\parallel}{4\times10^{28}\cm^2/\mathrm{s}}\right)^\frac12
\left(\frac{B}{5\mkG}\right)^{-\frac12}
\left(\frac{\ell}{10\p}\right)^{-\frac12}\!.
\end{equation}
In \Fig{ptau} we
illustrate the one-dimensional spread
of an initial Gaussian distribution of cosmic rays,
$e_{\rm c}=\exp(-\half x^2/\ell^2)$ after $t=\tau$ for three
values of $\mbox{St}$.
For small values of $\mathrm{St}$, the solution evolves similarly to that of
the diffusion equation (solid and dotted lines in \Fig{ptau}).
For large values of $\mathrm{St}$,
the distribution of cosmic rays develops two local maxima of $e_\mathrm{c}$
that propagate outwards as shown with dashed line, a typical wave-like behaviour.
In the limiting case of very large values of $\mathrm{St}$ the governing
equation reduces to the wave equation,
and the classical diffusion is recovered for $\mathrm{St}\to0$.

In some sense,
the extra time derivative in the non-Fickian formulation
plays a role similar to that of the displacement current in electrodynamics.
In simulations of hydromagnetic flows at low density, where the
Alfv\'en speed can be very large, the displacement current can be
included with an artificially reduced value of the speed of light
in order to limit the Alfv\'en speed to numerically acceptable values
\citep{MillerStone00}.

A comment regarding centred finite difference schemes is here in order.
In the steady state, the discretization of the cosmic ray diffusion
model given by
\Eqs{FluxDivergence}{nonFickian} corresponds essentially to a conservative
formulation of the diffusion term.
(A conservative formulation involving a direct discretization of
$\nabla^2$ is not possible with a non-staggered mesh, because two
first-order derivatives occur in two separate equations.)
As is well known, the discretization of the diffusion term on a centred
non-staggered mesh means that structures at the mesh scale
cannot be diffused (the discretization error for first derivatives
becomes infinite).
Therefore, we need to include weak
Fickian diffusion in the cosmic ray energy equation.
We refer to the corresponding (isotropic) diffusion coefficient
as $K_{\rm Fick}$, and it will be chosen to be comparable to or less than
the viscous and magnetic diffusivities.

In the following we use the \textsc{Pencil Code},\footnote{
\url{http://www.nordita.dk/software/pencil-code}}
a non-conservative,
high-order, finite-difference code (sixth order in space and
third order in time) for solving the compressible hydromagnetic equations.
The non-Fickian diffusion formulation is invoked by using the
\texttt{cosmicrayflux} module.
Whenever possible we display the results in non-dimensional form,
normalizing in terms of physically relevant quantities.
In all other cases we display the results in code units, which means that
velocities are given in units of the sound speed $c_{\rm s}$,
length is given in units of $k_1^{-1}$
(related to the scale of the box),
density is given in units of the average density $\rho_0$,
and magnetic field is given in units of $\sqrt{\mu_0\rho_0}\,c_{\rm s}$.
The units of all other quantities can be worked out from these.
For example, the unit of $Q_{\rm c}$ is $\rho_0 c_{\rm s}^3k_1$.
For the interstellar medium with
$\rho_0=10^{-24}\g\cm^{-3}$, $c_{\rm s}=10\km\s^{-1}$, and $k_1=2\pi/100\p$,
the unit of the cosmic ray injection rate is $3\times10^{-26}\erg\cm^{-3}\s^{-1}$, which
is about 10\% of the rate of energy injection by
supernovae in the galactic disc \citep{MLK04}.
The unit for diffusivity is
$c_{\rm s}k_1^{-1}\approx5\times10^{25}\cm^2\s^{-1}$.

\subsection{Cosmic ray diffusion near a magnetic {\sf X}-point}
\label{Xpoint}

\begin{figure}\begin{center}
\includegraphics[width=.41\textwidth]{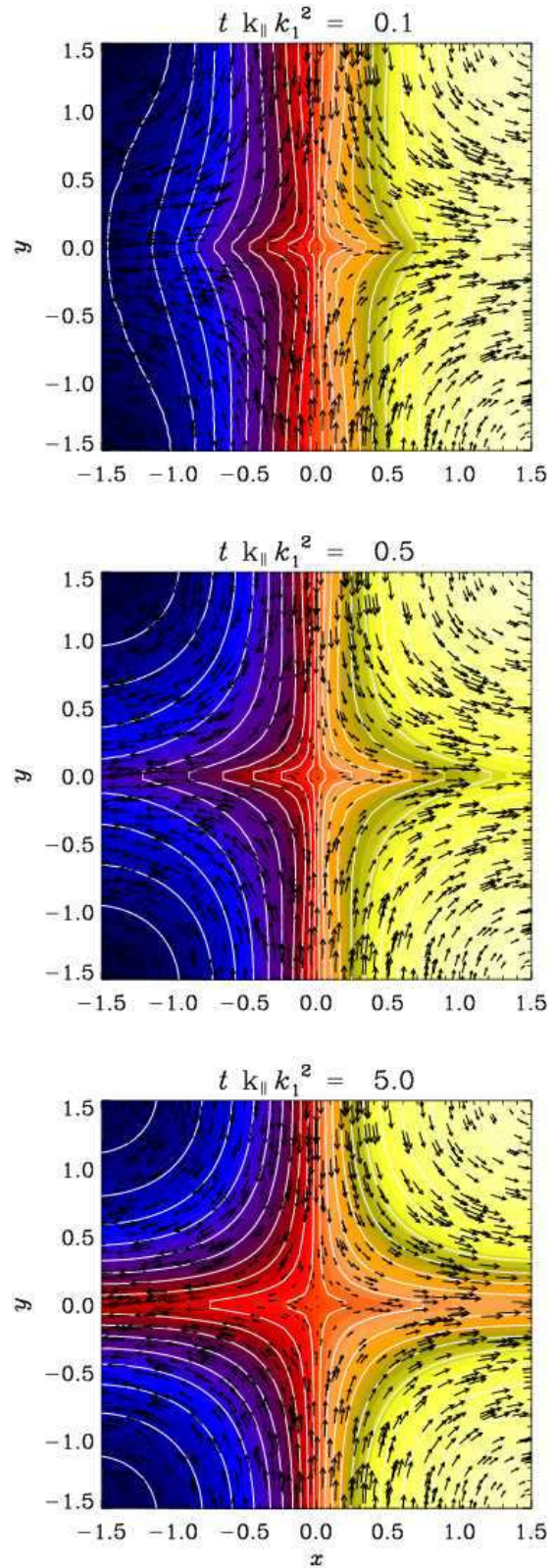}
\end{center}\caption[]{
Evolution of the cosmic ray energy density near a magnetic ${\sf X}$-point:
snapshots of $e_{\rm c}$ (shown as contours and shades of grey/colour) for field-aligned
diffusion along a fixed magnetic field $\BB=(\sin k_1x,-\sin k_1y,0)^T$
(shown as vectors) displayed for three times indicated at the top of each frame.
}\label{fXpoint}\end{figure}
We test the field-aligned diffusion procedure by simulating
in two dimensions
a magnetic field configuration similar to the {\sf X}-point discussed in
Sect.~\ref{NonFickian}.
In order to be able to impose normal-field boundary conditions,
$\nnn\times\BB=0$ at the domain boundaries, we modify the field to $\BB=(\sin k_1x,-\sin k_1y,0)^T$,
where $k_1$ is the smallest wavenumber in a periodic domain.
So, for $k_1=1$ we consider the domain $-\pi<(x,y)<\pi$.
The initial distribution of the cosmic ray energy density
is $e_{\rm c}=x$, which has a constant gradient and therefore,
with Fickian diffusion,
$D_c=\nab\cdot(\BBhat\BBhat)$ would have a singularity initially.
However, in the non-Fickian approach $D_{\rm c}$ is not calculated
as in \Eq{Dc}, which resolves this problem.
The evolution of $e_{\rm c}$ for $\tau = 0.1$ is shown in \Fig{fXpoint} together
with vectors showing the magnetic field.
Note that the gradient of $e_{\rm c}$ becomes small
in the neighbourhood of the singularity of $\nab\cdot(\BBhat\BBhat)$
at the origin, so the otherwise singular term that multiplies
$\nab e_{\rm c}$ has no effect on $e_\mathrm{c}$, as desired.
In the case of the Fickian diffusion, the same final solution would have been obtained,
but the initial reduction of the gradient in $e_\mathrm{c}$ would have
involved an infinitely large advection speed $\UU_{\rm c}$.
In the non-Fickian approach, the maximum propagation speed is
$\tilde{K}_\parallel^{1/2}$, thereby alleviating the numerical time step problem.

\begin{figure}\centering
\includegraphics[width=0.46\textwidth]{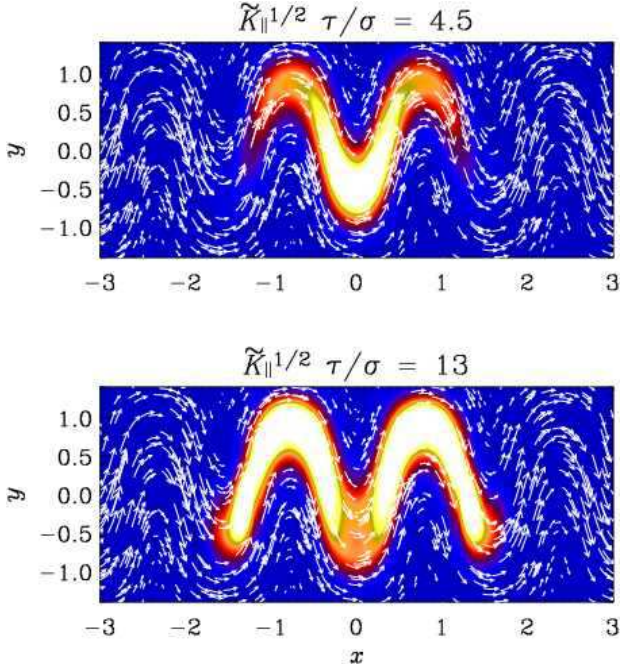}
\caption{
Magnetic field vectors together with a
grey/colour scale representation of $e_{\rm c}$
in a kinematic calculation with $128^2$ mesh points,
for different values of $K_\parallel^{1/2}\tau/\sigma$,
with $\tilde{K}_\perp=0$, $\tilde{K}_\parallel=10^{-1}$,
and $K_{\rm Fick}=10^{-3}$, at time $t/\tau=1$
for two different values of $\tau$ (=1 and 3, respectively).
(Only part of the computational domain in the $y$ direction is shown.)
}\label{pslice_all}\end{figure}

Another example of field-aligned diffusion is shown in \Fig{pslice_all},
where the magnetic field is given by $\BB=\BB_0+\nab\times\AAA$
with $\BB_0=0.1\xxx$ and $\AAA=0.1\zz\cos(k_xx)\cos(k_yy)$
with $k_x=4k_1$ and $k_y=k_1$.
Again, this magnetic field is held constant in time.
The initial profile of $e_\mathrm{c}\propto\exp(-r^2/2\sigma^2)$, with $r^2=x^2+(y+0.5)^2$,
is a two-dimensional Gaussian of a half-width
of $\sigma=0.07$, positioned at $(0,-0.5)$.
We confirm that our implementation of cosmic ray
diffusion allows us to model reliably rather complicated magnetic configurations.
The lower panel of \Fig{pslice_all} confirms that, for large values of the
Strouhal number, the wave nature of the telegraph equation manifests itself
and $e_\mathrm{c}$ develops two waves
propagating away from the initial maximum (similar to
the dashed line in \Fig{ptau}).

\section{Macroscopic evolution of the cosmic ray gas}

\subsection{Energy considerations}

In a closed domain, mass is conserved, i.e.\ $\bra{\rho}\equiv\rho_0=1$,
where angular brackets denote volume averaging. Then
the hydromagnetic equations coupled
with cosmic ray dynamics lead to the following set of
equations for the cosmic ray energy $E_{\rm c}=\bra{e_{\rm c}}$,
the gas energy $E_{\rm g}=\bra{e_{\rm g}}$,
the kinetic energy $E_{\rm k}=\bra{\half\rho\uu^2}$,
and magnetic energy $E_{\rm m}=\bra{\BB^2}/2\mu_0$,
\EQ
{\dd E_{\rm c}\over\dd t}=-W_{\rm c}+\bra{Q_{\rm c}},
\label{Ec}
\EN
\EQ
{\dd E_{\rm g}\over\dd t}=-W_{\rm g}+\bra{Q_{\rm k}}+\bra{Q_{\rm m}},
\label{Eg}
\EN
\EQ
{\dd E_{\rm k}\over\dd t}=W_{\rm c}+W_{\rm g}+W_{\rm m}+W_{\rm f}-\bra{Q_{\rm k}},
\label{Ek}
\EN
\EQ
{\dd E_{\rm m}\over\dd t}=-W_{\rm m}-\bra{Q_{\rm m}}.
\label{Em}
\EN
Here, all the energies are referred to the unit volume.
The terms $W_{\rm c}=\bra{p_{\rm c}\nab\cdot\uu}$,
$W_{\rm g}=\bra{p_{\rm g}\nab\cdot\uu}$,
$W_{\rm m}=\bra{\uu\cdot(\JJ\times\BB)}$, and
$W_{\rm f}=\bra{\uu\cdot\ff}$ result from
work done against cosmic ray pressure, gas pressure,
the Lorentz force, and the external forcing, respectively.
Terms responsible for viscous and Joule heating and the cosmic ray
energy source are simply given by the volume integrated terms
in the original equations.
Equations \eq{Ec}--\eq{Em} imply that the total energy,
$E_{\rm tot}=E_{\rm c}+E_{\rm g}+E_{\rm k}+E_{\rm m}$,
satisfies the simple conservation law
\EQ
{\dd E_{\rm tot}\over\dd t}=\bra{Q_{\rm c}}+W_{\rm f}.
\EN
Thus, the only sources of energy are the injection of cosmic rays
and the external forcing of the turbulence.
In the following section we demonstrate how $E_{\rm c}$ can be
enhanced by the conversion of kinetic energy.

\subsection{Compressional enhancement of cosmic ray energy}

\begin{figure}\centering\includegraphics[width=0.46\textwidth]{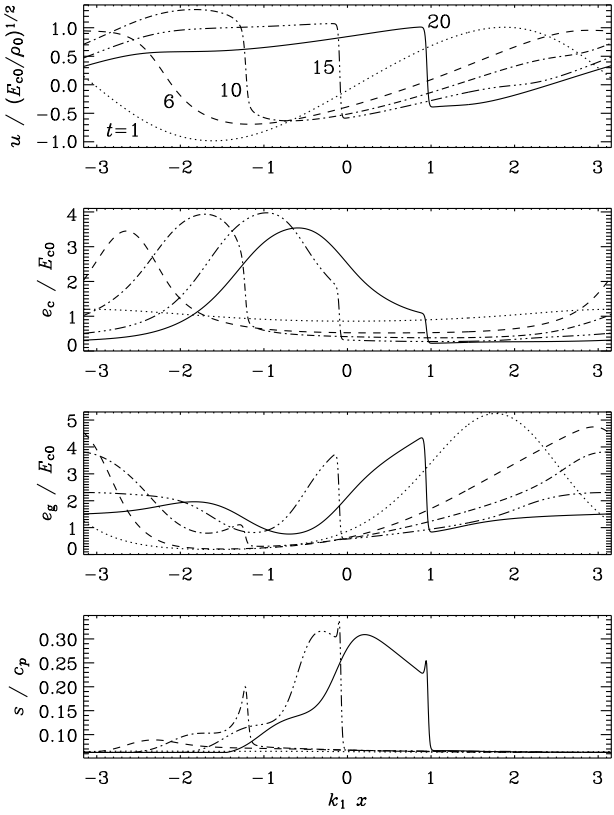}
\caption{
Velocity, cosmic ray and gas energy densities, and entropy in an experiment with
a non-linear sound wave that piles up to a shock ($\gamma_{\rm c}=5/3$).
Note the significant conversion of kinetic energy into cosmic ray energy.
The conversion into gas energy is comparatively small even though there is
noticeable entropy enhancement due to the shock.
Curves obtained for different times are shown with different line types as labelled
in the first panel.
Time is given in units of $k_1^{-1}(E_{\rm c0}/\rho_0)^{-1/2}$.
}\label{ptshock}\end{figure}

We assume $Q_{\rm c}=W_{\rm f}=0$ and that there is
initially kinetic energy that is later redistributed among gas and cosmic
rays.
We investigate, using a simple one-dimensional model
($\partial/\partial y=\partial/\partial z=0$), how much energy
can be converted into cosmic ray energy via the $W_{\rm c}$ term
responsible for work done against cosmic ray pressure.
As the initial condition, we use a sinusoidal perturbation of $u_x$ and
$\ln\rho$ with unit amplitude and $E_{\rm c}=E_{\rm c0}=1$,
$E_{\rm g}=1.8$, and $E_{\rm k}=0.21$.
The evolution of velocity, cosmic ray and gas energies, as well as the
entropy of the gas are shown in \Fig{ptshock}.
Here the entropy $s$ is defined as
$s=c_v\ln(c_{\rm s}^2/\rho_{\rm g}^{\gamma-1})$,
where $c_{\rm s}^2=\gamma(\gamma-1)e_{\rm g}$ is the gas sound speed squared.
It turns out that in this case about 78\% of the kinetic energy
is transformed into cosmic ray energy and only 22\% into thermal energy.
This result is however sensitive to the phase
shift between density and velocity: if the
density is initially uniform (keeping all other parameters unchanged),
the fractional energy going into cosmic rays is only 23\% and 77\%
go into thermal energy.

These results demonstrate that, at least in principle, a sizable fraction
of the kinetic energy can be converted into cosmic ray energy.
Similar experiments have been made in earlier work with a similar model in the
context of shock acceleration of cosmic rays \cite[see, e.g.,][]{DV81,JCN94}.
In particular \cite{KJ90} showed that the efficiency of conversion varies
strongly with $\gamma_{\rm c}$.
However, the conversion of kinetic energy into cosmic ray energy
requires a background of cosmic ray energy.
Decreasing $E_{\rm c}$ from 1 to 0.1 lowers the fraction of compressionally
produced cosmic ray energy density from 78\% to 21\%.
In contrast to dynamo theory where a weak seed magnetic field is
sufficient to produce equipartition magnetic fields (albeit only in three
dimensions), there is no such mechanism for the cosmic ray energy.
This is related to the anti-dynamo theorem for scalar fields
\citep{Krause1972}.
However, for three-dimensional compressible flows an exponential
dynamo-like amplification of a passive scalar is in principle possible
if the passive scalar is represented by inertial particles \citep{EKR96}.
Such a mechanism can work because inertial particles do not feel a
pressure gradient.
This can lead to particle accumulation in temperature minima \citep{EKR97}
and in vortices \citep{BS95,HB98,JAB04}.
However, in this paper cosmic ray particles are treated as non-inertial
particles.

\subsection{Effect of cosmic ray pressure}
\label{CRpressure}

\begin{figure}\centering
\includegraphics[width=0.47\textwidth]{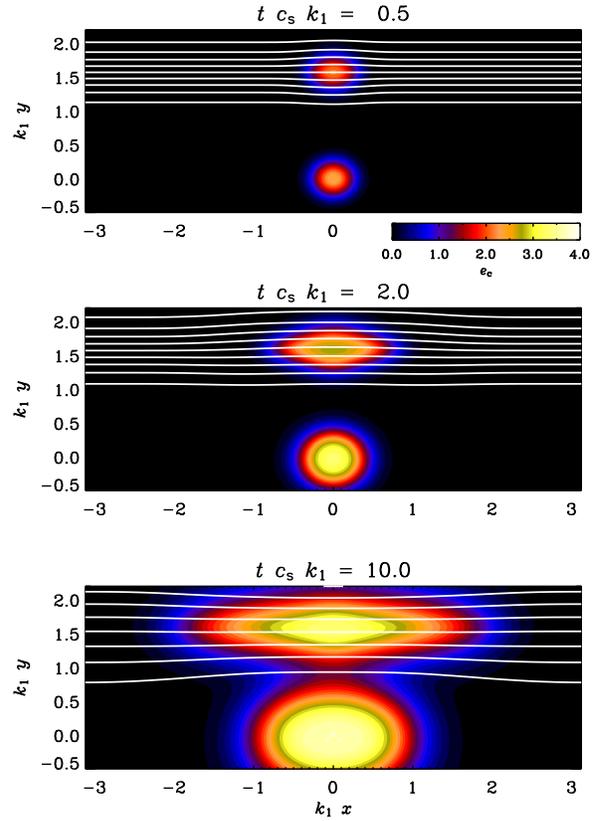}
\caption{
Cosmic ray energy density at times
indicated at the top of each panel.
Cosmic rays expand from two sources (with injection rate
$Q_{\rm c}=10$ for each), one inside a magnetic flux tube
and the other one outside.
Magnetic lines are shown with white solid curves whose density
is proportional to the field strength.
}\label{pblowout_all}\end{figure}

Cosmic rays can be confined at large scales by magnetic tension, where
a strong magnetic field can more easily withstand
deformation driven by cosmic ray pressure gradients.
This could provide a natural mechanism for producing equipartition between
cosmic rays and the magnetic field.
This feature can be simulated in two dimensions in a doubly periodic domain
$-\pi<(x,y)<\pi$, with $k_1=1$.
The results are illustrated in \Fig{pblowout_all},
where we have a magnetic tube in $1<y<2$ with its axis along the $x$ direction.
We have implemented two local cosmic ray sources with
energy injection profiles
\EQ
Q_{\rm c}=Q_{\rm c0}
\sum_{i=1}^2\exp\left\{-\half\left[x^2+(y-y_i)^2\right]/R^2\right\},
\EN
i.e., both located on the $y$ axis, centred at $y_1=0$ and $y_2=\pi/2$;
the initial half-widths for both sources is $R=0.13$, so that one source
is within the magnetic tube and the other, outside it.
In this experiment, cosmic ray diffusion is negligible
($\tilde{K}_\parallel=\tilde{K}_\perp=0$ and $K_{\rm Fick}=0.01$)
as we intend to explore the effects of cosmic ray pressure alone.
As expected, expansion proceeds nearly
isotropically outside the magnetic structure,
but the cosmic ray energy density is
channelled preferentially along field lines inside the tube.
At the end of the run, the aspect ratio
of the cosmic ray distribution is about two to one inside the tube.
For values of $Q_{\rm c}$ significantly larger than
about 10, the gas density decreases strongly so as to maintain pressure
equilibrium and oppose expansion driven by cosmic rays.

This confirms that cosmic ray dynamics can be strongly affected by the
approximate pressure balance in the ISM.

\subsection{Cosmic rays in a partially ordered magnetic field}
\label{RandomField}

In this section we briefly explore the effects of a random magnetic field
on the evolution of the cosmic ray gas. A random component of the interstellar
magnetic field can facilitate the isotropic spreading of cosmic rays across the large-scale,
preferentially horizontal magnetic field in the Galactic disc.
In addition, a turbulent magnetic field can enhance cosmic ray
diffusion by destroying the compound diffusion effect \citep[][and references therein]{P79,KJ00}
due to the exponential local divergence of magnetic lines.

To allow for cosmic ray losses through the
$x$ boundaries, we relax the assumption of periodicity
in that direction.
At $x=\pm \pi$, we assume $e_{\rm c}=0$, together with
$\partial\rho/\partial x=\partial e_{\rm g}/\partial x=0$.
This implies that cosmic rays may be lost from the domain
but gas may not.
In the $y$ direction we again use periodic boundary conditions.

\begin{figure}\centering
\includegraphics[width=0.46\textwidth]{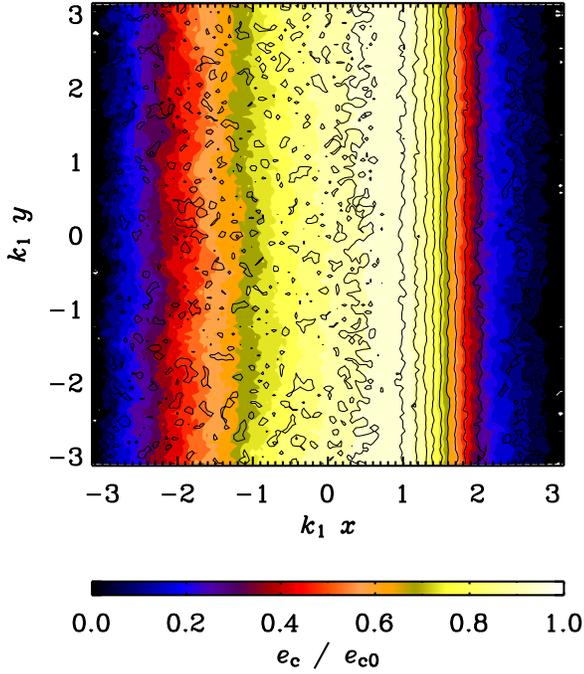}
\caption{
Cosmic ray energy density (colour/grey scale coded, with darker/blue shades
corresponding to smaller values)
together with magnetic field lines (solid) in a two-dimensional simulation
with a fixed magnetic flux tube centred around $x=1.5$ and
a random magnetic field superimposed on it.
Here, $\tilde{K}_\parallel=0.1$, $\tilde{K}_\perp=0$, and $\tau=3$.
}\label{emb_flux2d}\end{figure}

\begin{figure}\centering
\includegraphics[width=0.46\textwidth]{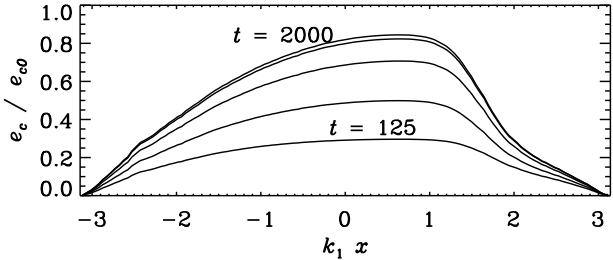}
\caption{
Cosmic ray energy density from the model of \Fig{emb_flux2d} averaged in the $y$ direction
for times $125 \times 2^n$ with $n=0,...,4$.
The magnetic tube is located at $x=1.5$ leading to an
asymmetric distribution of cosmic ray energy density.
}\label{pecmxt}\end{figure}

We consider a two-dimensional system with a regular magnetic field $\BB_0$
directed along the $y$-axis and confined to a flux tube
as shown in \Fig{emb_flux2d}, where the
field strength has a profile $B_0\propto\mbox{sech}^{2}[(x-1.5)/0.5]$.
An isotropic random magnetic field $\delta\BB$ is superimposed on $\BB_0$, with $\mean{\delta B^2}/B_0^2=1$
at $x=1.5$ where $B_0$ is maximum; the magnetic field does not evolve.
The random magnetic field is implemented in terms of a magnetic vector potential
given as white noise with Gaussian probability density which,
because of two dimensions, implies a $k^3$ power
spectrum for the magnetic energy.
We also assume zero velocity for all times, so we just advance
\Eqs{decdt}{nonFickian} in time, using \Eq{FluxDivergence}.
Cosmic rays are injected at a constant rate across the domain,
$Q_\mathrm{c}=\mathrm{const}$.

In \Fig{emb_flux2d} we show the result of such a calculation with $\tilde{K}_\perp = 0$ ;
the distribution of cosmic rays in $x$ is asymmetric reflecting the asymmetry
in the relative amount of disorder of the magnetic field, $\mean{\delta B^2}/B_0^2$.
This asymmetry can be seen more clearly
in \Fig{pecmxt} which shows the evolution of cosmic ray energy density averaged
in the $y$-direction.
(Note however that the steady state is only attained after very long times.
Here, $t=2000$ corresponds to $t \tau\tilde{K}_\parallel k_1^2=600$.)

\begin{figure}\centering
\includegraphics[width=0.45\textwidth]{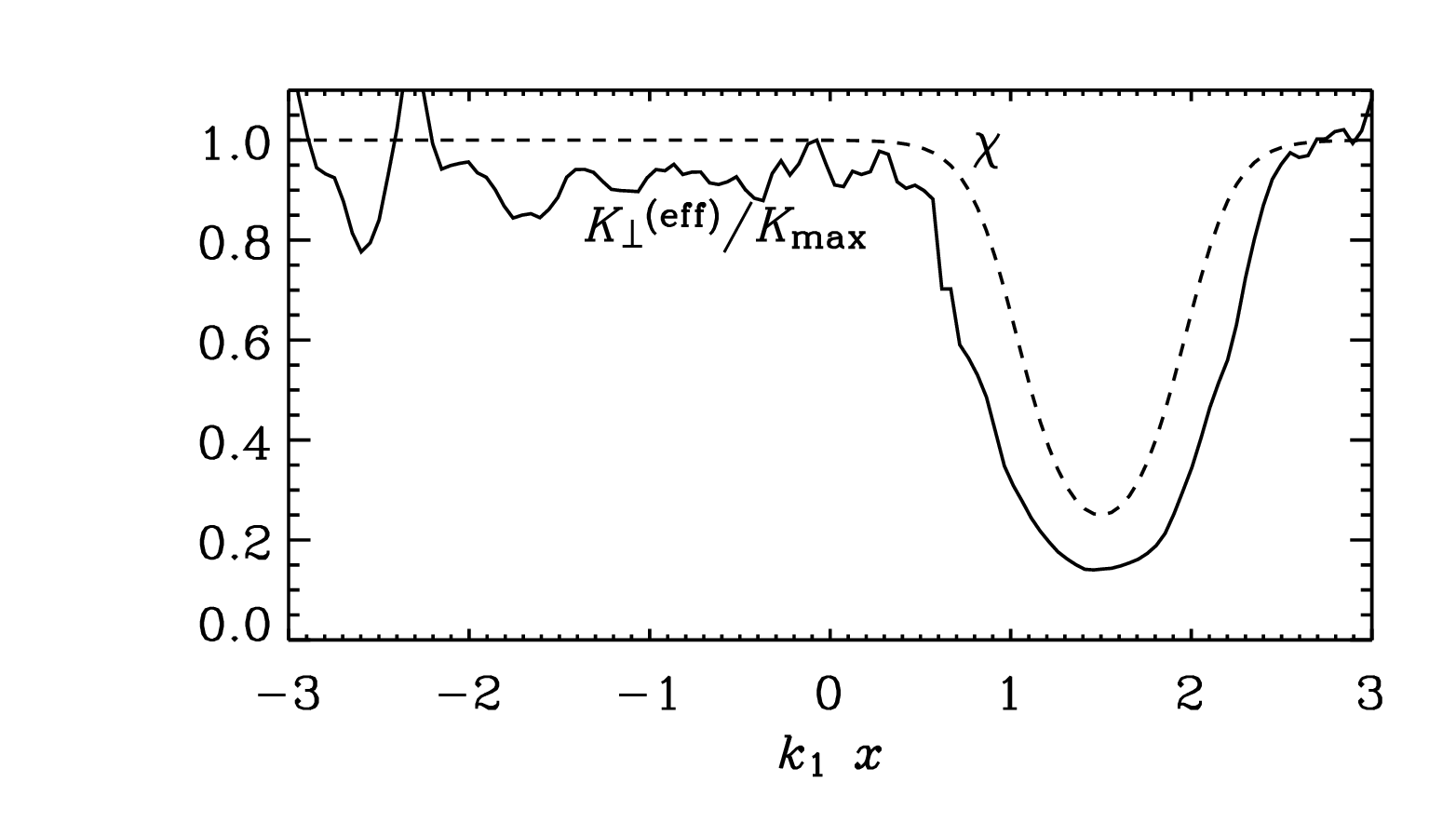}
\caption{
The profile of $K_\perp^\mathrm{(eff)}$  (solid) obtained from \Eq{KeffSolution}
using $e_{\rm c}$  corresponding to the upper curve of \Fig{pecmxt},
and $\chi=3\mean{B_x^2}/\mean{B_y^2}$ (dashed), where $B_y$ has both large-scale
and random parts, whereas $B_x$ is a purely random magnetic field.
Here, $K_\mathrm{max}=0.023$ is the maximum value of $K_\perp^\mathrm{(eff)}$.
}\label{keff}\end{figure}

The effective perpendicular diffusivity due to the
randomness of the magnetic field, $K_\perp^\mathrm{(eff)}(x)$,
can be obtained from the steady-state equation
\EQ
{\dd\over\dd x}\left(K_\perp^\mathrm{(eff)}(x){\dd e_\mathrm{c}\over\dd x}\right) = -Q_\mathrm{c},
\label{KeffEquation}
\EN
which can be integrated to obtain
\EQ
K_\perp^\mathrm{(eff)}(x)=
(x_0 - x) Q_\mathrm{c}\left({\dd e_{\rm c}\over\dd x}\right)^{-1},
\label{KeffSolution}
\EN
where $x_0$ is the position where $\dd e_\mathrm{c}/\dd x=0$.
The resulting profile of $K_\perp^\mathrm{(eff)}$, shown in \Fig{keff}
along with
\EQ\label{chi}
\chi=\frac{\mean{\delta B^2}}{B_0^2+{\textstyle{1\over3}}\delta B^2}
=3\,\frac{\mean{B_x^2}}{\mean{B_y^2}},
\EN
confirms that
the effective perpendicular diffusion is controlled by the degree of randomness
of the magnetic field \citep[see, e.g.,][]{CP93}.

\section{Cosmic rays in a magnetic field produced by dynamo action}
\label{Turbulence}

Three-dimensional turbulence is capable of dynamo action
for sufficiently large magnetic Reynolds numbers, and the dynamo-generated magnetic
field organizes itself into random flux tubes
or sheets \citep[e.g.][ and references therein]{Z90,BPS95,BS05}.
Most studies of cosmic ray dynamics neglect the specific features of the
magnetic fields produced by turbulent dynamos. We provide here a preliminary
discussion of cosmic ray evolution in a magnetic field generated by
a turbulent flow of electrically conducting fluid.
The magnetic field structure of these simulations is realistic enough to
include important physical effects such as the enhancement of cosmic ray
diffusion by turbulent fields, as mentioned in \Sec{RandomField}.

Magnetic field produced by the dynamo action is rather different from that
prescribed as, say, a random vector field with given spectrum
and Gaussian statistical properties of the components. In contrast to such
{\em ad hoc\/} models, dynamo magnetic fields can be
strongly intermittent (i.e., dominated by intense magnetic filaments, ribbons
and sheets) and varying in time \citep[see][ and references therein]{BS05}; both
features can affect the propagation of charged particles. Moreover, since
both gas flow and magnetic
field are random (in space and time), any relation between cosmic ray
energy density and other parameters of the medium (e.g., magnetic energy
density or gas density) can only be statistical. Therefore, we expect that
the energy density of cosmic rays can locally (and at any given moment)
significantly exceed, say, the magnetic energy density. However, one would
expect that some form of equipartition between energy densities (or forces
due to) cosmic rays and magnetic fields can be maintained {\em on  average.}
We note, however, that simulations have not fully
confirmed these expectations; see also \citet{PadoanScalo05}.

Our model is realistic with respect to modelling fully nonlinear dynamo
action as we simulate consistently both a randomly forced flow and the magnetic field
produced by it, by solving both the Navier--Stokes and induction equations
(with the Lorentz force included in the former, and the velocity field obtained
from the Navier--Stokes equation in the latter).
The turbulent motions in our model are driven by a random force explicitly
included in the Navier--Stokes equation.
In reality, interstellar turbulence is driven by
supernova explosions that produce strongly compressible
flows with very large local Mach numbers locally \citep[some aspects of the
relevant models are reviewed by][]{MLK04}. However, we deliberately restrain ourselves from a
detailed discussion of such more realistic models here (which would also
include stratification, disc--halo connections, velocity shear, etc.), but
instead explore just the effects of magnetic intermittency and variability. We
believe that our simulations capture at least some of the most important
effects of interstellar dynamo action on the cosmic ray propagation (within the limits
of our model for the cosmic rays).

The turbulence in our simulations is driven helically by a forcing function
$\ff$ in the Navier--Stokes equation, as was done in the simulations of
\cite{B01}, for example.
At $x=\pm\pi$, we use stress-free normal field boundary conditions
\citep[as was also done in][]{BD01}, and assume $e_{\rm c}=0$ on the boundaries as in
\Sec{RandomField}.
In the other directions we take periodic boundary conditions.
Our analysis of the results presented below only uses positions that are some distance
away from the domain boundaries ($L_x/8$ on both boundaries)
to reduce their influence.
(Including boundary points merely tends to decrease the magnitude of the
correlation coefficients between the various energy densities, but it does 
not change the results qualitatively.)
The forcing function is given in \App{ForcingFunction} and
its (dimensionless) amplitude for the simulation shown here is chosen to be $f_0=2$,
which produces an rms Mach number of about 1.2.

\begin{figure}\centering
\includegraphics[width=0.46\textwidth]{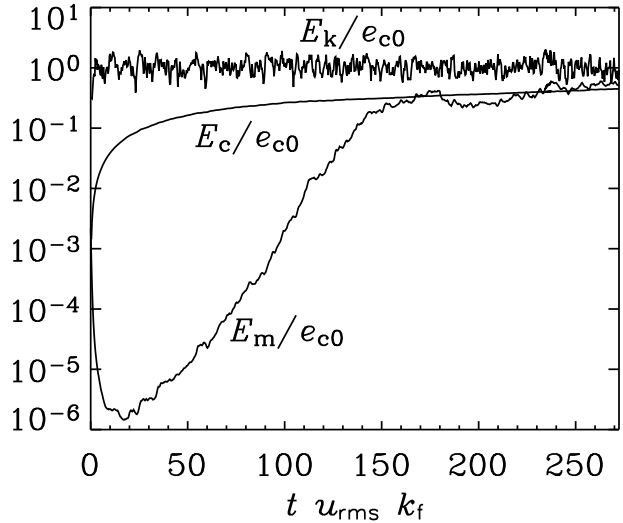}
\caption{
Time series of magnetic ($E_\mathrm{m}$), kinetic ($E_\mathrm{k}$) and cosmic ray
($E_\mathrm{c}$) energies
in a dynamo simulation.
Here, time is given in turnover times $(u_{\rm rms}k_{\rm f})^{-1}$, and
$e_{\rm c0}=L_x^2 Q_{\rm c}/K_\parallel$ is used to normalize
energies per unit volume.
The thermal energy of the gas is constant with
$E_{\rm g}/e_{\rm c0}\approx0.7$.
}\label{ptime}\end{figure}

The forcing wavenumber is chosen to be $k_{\rm f}=1.5\,k_1$.
This value is close to the wavenumber corresponding to the box size, $k_1=2\pi/L_x$,
so we do not expect to have clearly distinct large-scale and small-scale magnetic fields.
Generally, the flow helicity allows us to obtain dynamo action at relatively
small values of the magnetic Reynolds number defined as
$R_{\rm m}= u_{\rm rms}/(\eta k_{\rm f})$.
However, because of the non-periodic boundaries in the $x$ direction, and
also because of the weak scale separation
($k_{\rm f}/k_1$ is not very large), the critical value
of $R_{\rm m}$
with respect to the onset of dynamo action is still around
${\Rm}_\mathrm{,cr}=30$, which is similar to what would be expected
for a non-helical random flow in a periodic domain.
The simulation presented here has $R_{\rm m}\approx150$.
The kinematic growth rate
of the rms magnetic field is about $0.06\,u_{\rm rms}k_{\rm f}$.
In \Fig{ptime} we show the evolution of the magnetic energy together
with kinetic and cosmic ray energies.
We see that the magnetic field grows exponentially for
$t\la150/(u_{\rm rms}k_{\rm f})$ and then saturates---in agreement
with earlier simulations quoted above.
We note that the energy density of cosmic rays is much larger than magnetic
energy density at these early times; nevertheless, the cosmic ray energy
increases rather slowly after $t\ga50/(u_{\rm rms}k_{\rm f})$.
The steady-state energy density of cosmic rays is controlled by their
injection rate $Q_\mathrm{c}$ and their diffusivity:
solutions of Eq.~(\ref{KeffEquation}) are proportional to
$Q_\mathrm{c}/K^{(\rm eff)}_\perp$. However, the effective diffusivity of cosmic rays is
controlled by the degree of tangling of the magnetic field
rather than by the field strength itself; see, e.g., Eq.~(\ref{chi}).
It is not surprising, then, that even a weak magnetic field can confine
cosmic rays at early times in this model.
The linear dependence of the steady-state
energy density of cosmic rays on their injection rate is a direct consequence of the
(almost) linear nature of the cosmic ray
dynamics as described by Eq.~(\ref{decdt}); the only nonlinearity here is that the cosmic ray energy
density affects the flow through the pressure term, and then the velocity
field enters the induction equation and the advection term for the cosmic rays.
However, this nonlinearity is not very strong, and our simulations confirm a
linear dependence of $e_\mathrm{c}$ on $Q_\mathrm{c}$ within a broad range
of the latter (at least two orders of magnitude).
The magnetic field part $B_0$
is understood, in the present context, as an average over a scale
smaller than the domain size but larger than, say, the gyro radius of cosmic ray
particles.

For the simulation shown here
we have chosen
$Q_{\rm c}=0.01$, which yields
a steady-state cosmic ray energy of
$E_{\rm c}\approx1$ in units of
$L_x^2 Q_\mathrm{c}/K_\parallel$.
The other parameters of the simulation presented here are
$\tilde{K}_\perp=0$, $\tilde{K}_\parallel=0.3$, $K_{\rm Fick}=2\times10^{-2}$,
$\tau=0.3$, $\eta=5\times10^{-3}$, $\nu=0.5$.
Furthermore, because the Mach number is
slightly larger than unity,
an additional bulk
viscosity proportional to the negative velocity divergence has been
included.
This is usually referred to as a shock viscosity; see \cite{Haugen04} for
details and the definition of a non-dimensional parameter $c_{\rm shock}$
which is here chosen to be 10.
The value of $\tilde{K}_\parallel$ is chosen to be close to the maximum
Alfv\'en speed squared.
The magnetic field produced by the dynamo has pronounced magnetic filaments whose
half-width (radius) is about $\ell=0.2$, which is consistent with
the estimate $\ell\simeq \pi k_\mathrm{f}^{-1}R_{\rm m,cr}^{-1/2}$ suggested by
\citet{S99}.
For $\tau=0.3$ and $\ell=0.2$, we have $\mbox{St}\approx1$ from \Eq{Stdef}.
The steady-state mean kinetic energy density depends directly
on the intensity of the forcing. On the other hand, the ratio of magnetic to
kinetic energy densities is controlled by the nature of the dynamo action.
The above parameter values have been chosen as to ensure that the
energy densities of magnetic field and cosmic rays are of the same order of
magnitude in the statistically steady state.

\begin{figure}\centering
\includegraphics[width=0.46\textwidth]{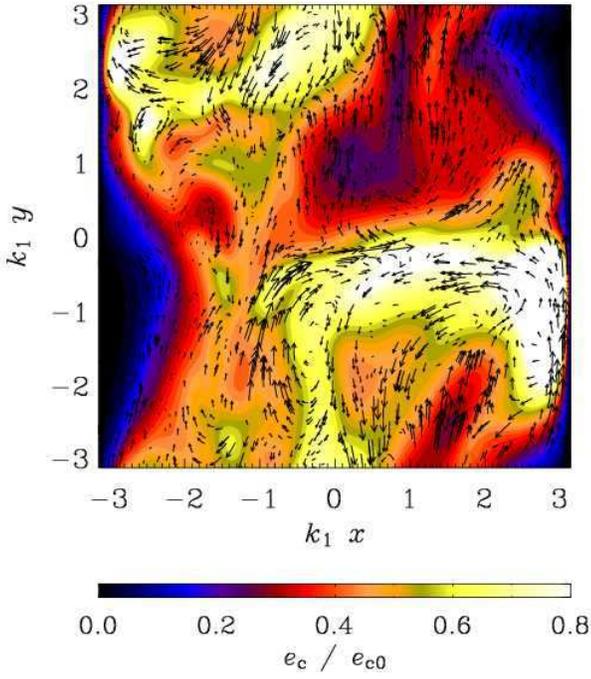}
\caption{
Cosmic ray energy density (colour/grey-scale coded, with lighter shades/redder
colour corresponding to larger values) and magnetic field
vectors in a slice taken from a dynamo simulation.
The magnetic field vectors are more numerous where magnetic field strength is
larger.
}\label{pbb_bc128c}\end{figure}

\begin{figure}\centering
\includegraphics[width=0.46\textwidth]{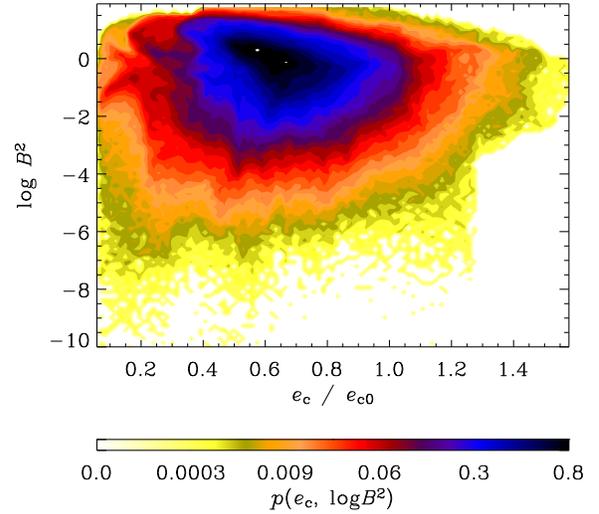}
\caption{
Two-dimensional histogram (or joint probability density)
of magnetic pressure and cosmic ray energy density.
Here, $e_{\rm c0}=L_x^2 Q_{\rm c}/K_\parallel$ is used
to normalize $e_{\rm c}$.
The two-dimensional probability density is calculated 
using only points at a distance greater than $L_x /8$ from the boundaries
in an attempt to avoid the regions where the distribution of
$e_\mathrm{c}$ is affected by the boundary conditions.
}\label{bc128c_Bpdf}\end{figure}

\begin{figure}\centering
\includegraphics[width=0.46\textwidth]{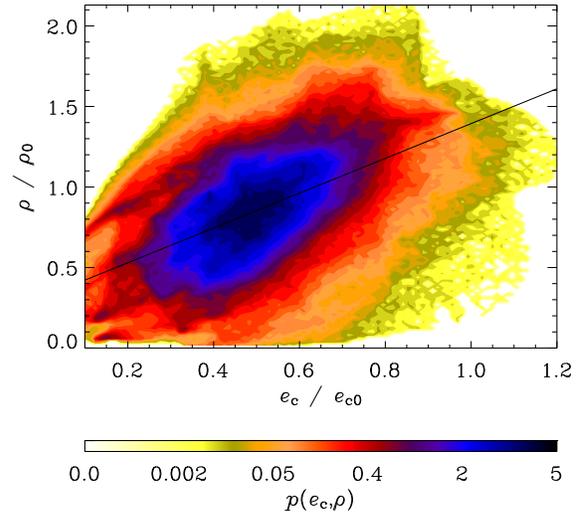}
\caption{
As in \Fig{bc128c_Bpdf}, but for
gas density and cosmic ray energy density, showing a modest
correlation between the two.
The correlation coefficient is $r=0.54$, and the straight line
is a best-fitting line.
}\label{bc128c_pdf}\end{figure}

In \Fig{pbb_bc128c} we show a typical cross-section of
the cosmic ray energy density and magnetic field vectors from the
three-dimensional dynamo simulation of \Fig{ptime} at
$t=250/(u_{\rm rms}k_{\rm f})$.
The cosmic ray energy density declines toward the
boundaries at $x=\pm\pi$, where the boundary condition $e_\mathrm{c}=0$ is imposed,
and shows some moderate variation inside
the domain. There is no pronounced correlation with magnetic field strength
even though imprints of the field-aligned diffusion can clearly be seen,
e.g., between $(x,y)k_1=(-1,-1)$ and $(0,0)$.
We show in \Fig{bc128c_Bpdf} a two-dimensional joint probability density
function
of $\log B^2$ and $e_\mathrm{c}$ (normalized to unit integral as usual), which
demonstrates the lack of any noticeable correlation between these variables.
The finite lifetime of magnetic structures produced by the dynamo must be
one of the reasons of the lack of correlation between the two variables.
There is some correlation between gas density and cosmic ray energy
density, as shown in \Fig{bc128c_pdf}, but
the cross-correlation coefficient is only 0.54, with the best-fitting dependence
$e_\mathrm{c}/e_\mathrm{c0}\simeq\rho/\rho_0$.

If the injection rate of cosmic rays is reduced by a factor of ten to
$Q_\mathrm{c}=10^{-3}$, the resulting steady-state mean value of the
cosmic ray energy density is found to be reduced by about the same factor.
The relation between cosmic ray energy density and gas density still appears to be
nearly linear, but the cross-correlation
coefficient is now larger, varying with time in the range 0.7--1.

We have also explored a run with a Mach number of 0.2 (achieved
by using a weaker driving force; $f_0=0.05$) but with the same injection rate of cosmic rays
as above, $Q_\mathrm{c}=0.01$.
The resulting steady-state mean energy density of cosmic rays exceeds those of
magnetic field and turbulence,
$E_\mathrm{c}/e_\mathrm{c0}\simeq1$, $E_\mathrm{k}/e_\mathrm{c0}\simeq0.02$
and $E_\mathrm{m}/e_\mathrm{c0}\simeq0.01$.
This produces significant anticorrelation between cosmic ray energy density
and gas density (cross-correlation coefficient of $-0.94$), with a linear dependence
between $e_\mathrm{c}$ and $\rho$.

The latter anticorrelation may be attributed to the average pressure equilibrium
in the domain, while a positive correlation in the supersonic flow may arise
as both gas and cosmic rays are compressed by the gas flow.
We have confirmed that no positive correlation between cosmic rays and gas
density occurs if the cosmic ray advection is neglected.

The model illustrated in \Figs{ptime}{pbb_bc128c} is close
to energy equipartition between cosmic rays, magnetic field and turbulence.
We note however that the Lorentz force and the cosmic ray pressure gradient
have very different magnitudes because the field-aligned cosmic ray
diffusivity is much larger than the magnetic diffusivity.
As a result, the cosmic rays are distributed
more uniformly than the magnetic field and the gas density and so
the cosmic ray pressure gradient
is comparatively small. For the values of the diffusivities given above,
the ratio the rms cosmic ray pressure gradient, $F_\mathrm{c}$,
and the rms Lorentz force $F_\mathrm{m}$, is typically
about $0.1$ of the ratio of the corresponding mean energy densities; this also applies
if the Lorentz force is replaced by the gradient of kinetic energy density.
The typical length scale of the magnetic field is about
$\ell\simeq l_0R_\mathrm{m,cr}^{-1/2}$ and $F_\mathrm{m}\simeq e_\mathrm{m}/\ell$,
with $l_0\simeq100\p$ the turbulent
scale and ${\Rm}_\mathrm{,cr}\approx30$ the critical magnetic Reynolds number
for the onset of dynamo action (see above).
The length
scale of the cosmic ray distribution can be estimated as the diffusion scale
over the confinement time $\tau_\mathrm{c}\simeq10^7\yr$,
$l_\mathrm{c}\simeq(K_\parallel\tau_\mathrm{c})^{1/2}$. Then, for
$K_\parallel=10^{28}\cm^2\s^{-1}$,
\[
\frac{F_\mathrm{c}}{F_\mathrm{m}}\simeq
\frac{\ell}{l_\mathrm{c}}\,\frac{E_\mathrm{c}}{E_\mathrm{m}}
\simeq\frac{1}{30}\,\frac{E_\mathrm{c}}{E_\mathrm{m}}\;.
\]
This conclusion appears to be
model-independent and suggests that energy equipartition between cosmic rays
and other constituents of the interstellar medium does {\it not\/} necessarily
imply that cosmic rays play an important role in the dynamical balance.

\section{Discussion and conclusions}

We have presented a preliminary analysis of cosmic ray propagation in a
magnetic field produced by dynamo action of a turbulent flow.
The confinement of cosmic rays resulting from their scattering by
magnetohydrodynamic
waves can be modelled with an equation similar to Eq.~(\ref{decdt}), where the
advection velocity is a linear combination of gas velocity and
Alfv\'en velocity \citep[][]{Skilling}. Our results are based on advection
with the local gas velocity.
\cite{PadoanScalo05} considered local variations in cosmic ray
density in the case  where the advection velocity
is given by the Alfv\'en velocity. They predict that
$e_\mathrm{c}\propto n_\mathrm{i}^{1/2}$, with $n_\mathrm{i}$ the ion density.
This scaling
is expected if the diffusive streaming velocity, $-K_\parallel \nabla
e_\mathrm{c}$, and the effects of cosmic ray pressure are negligible. Our model
can be adapted to test and generalize these results; the anticorrelation
between $e_\mathrm{c}$ and gas density in
one of our models (with low Mach number)
seems to be a direct consequence of pressure balance,
while a positive correlation (obtained at larger Mach number)
may reflect the fact that both cosmic rays and thermal gas experience
similar compression by the gas flow.
We have shown that our model captures naturally the dependence of the
effective diffusivity of cosmic rays on the ratio of random to ordered magnetic
field, $\delta B^2/B_0^2$.

The diffusivity of cosmic rays along the magnetic field is rather large; the
corresponding Strouhal number, defined in Eq.~(\ref{Stdef}) may significantly
exceed unity, as shown in Eq.~(\ref{Stest}). For
comparison, a similar estimate yields $\mathrm{St}\simeq1$ for the turbulent
kinetic and magnetic diffusivities in the ISM. This motivates our suggestion that the standard
Fickian diffusion model, which leads to the classical diffusion equation, may
be a poor approximation for cosmic rays, and a more accurate description leading to some form
of the telegraph equation might be more appropriate. Formally, the difference
between the two approximations consists of retaining, in the latter
approximation, higher-order terms in the correlation time of the random
process underlying diffusion. We have introduced this effect to alleviate
numerical problems, but it can be a real physical effect which deserves
further careful study.

In summary, we have found that the cosmic ray distribution can be
more uniform than the distributions of magnetic field and gas density.
Consequently, we may argue that energy equipartition between cosmic rays
and other constituents of the interstellar medium does not necessarily
imply that cosmic rays play a significant role in the dynamical balance.

\appendix
\section{Boundedness of cosmic ray energy density}
\label{Bounded}

In this section we show that, in a closed or periodic domain,
$\max(e_{\rm c})$ can only decrease as a result of (tensorial) diffusion.
This is useful for showing that the diverging behaviour of $U_{\rm c}$
does not produce a singularity in $e_{\rm c}$; cf.\ \Sec{NonFickian}.
In order to avoid interference from other effects, we assume that the
evolution of $e_{\rm c}$ is only governed by diffusion, i.e.\
\EQ
{\partial e_{\rm c}\over\partial t}=
\nabla_i\left(K_{ij}\nabla_je_{\rm c}\right).
\EN
Note also that $\max(e_{\rm c})=\bra{e_{\rm c}^n}^{1/n}$ for $n\to\infty$.
Here, angular brackets denote volume averages.
Thus, using integration by parts, we have
\EQA
{\dd\over\dd t}\bra{e_{\rm c}^n}
&=&n\bbra{e_{\rm c}^{n-1}{\partial e_{\rm c}\over\partial t}}
=n\bbra{e_{\rm c}^{n-1}\nabla_i\left(K_{ij}\nabla_j\right)}
\nonumber \\
&=&-n(n-1)\bbra{e_{\rm c}^{n-2}K_{ij}(\nabla_ie_{\rm c})(\nabla_je_{\rm c})}
\nonumber \\
&\leq&0\quad\mbox{(for any value of $n>1$)}.
\ENA
The last inequality assumes that the diffusion tensor is positive definite,
which is true in our case, because
\EQ
K_{ij}(\nabla_ie_{\rm c})(\nabla_je_{\rm c})\!=\!
(K_\parallel\!-\!K_\perp)(\BBhat\cdot\nab e_{\rm c})^2
\!+\!K_\perp(\nab e_{\rm c})^2
\EN
is positive.
Therefore, $\max(e_{\rm c})$ must decrease with time.

\section{The forcing function}
\label{ForcingFunction}

For completeness we specify here the forcing function used in the
present paper\footnote{This forcing function was also used by
Brandenburg (2001), but in his Eq.~(5) the factor 2 in the denominator
should have been replaced by $\sqrt{2}$ for a proper normalization.}.
It is defined as
\EQ
\ff(\xx,t)={\rm Re}\{N\ff_{\kk(t)}\exp[\ii\kk(t)\cdot\xx+\ii\phi(t)]\},
\EN
where $\xx$ is the position vector.
The wavevector $\kk(t)$ and the random phase
$-\pi<\phi(t)\le\pi$ change at every time step, so $\ff(\xx,t)$ is
$\delta$-correlated in time.
For the time-integrated forcing function to be independent
of the length of the time step $\delta t$, the normalization factor $N$
has to be proportional to $\delta t^{-1/2}$.
On dimensional grounds it is chosen to be
$N=f_0\rho_0 c_{\rm s}(|\kk|c_{\rm s}/\delta t)^{1/2}$, where $f_0$ is a
dimensionless forcing amplitude.
At each time step we select randomly one of many possible wavevectors
in a certain range around a given forcing wavenumber.
The average wavenumber is referred to as $k_{\rm f}$.
Two different wavenumber intervals are considered: $1$--$2$ for
$k_{\rm f}=1.5$ and $4.5$--$5.5$ for $k_{\rm f}=5$.
We force the system with transverse helical waves,
\begin{equation}
\ff_{\kk}=\RRRR\cdot\ff_{\kk}^{\rm(nohel)}\quad\mbox{with}\quad
{\sf R}_{ij}={\delta_{ij}-\ii\sigma\epsilon_{ijk}\hat{k}_k
\over\sqrt{1+\sigma^2}},
\end{equation}
where $\sigma=1$ for positive helicity of the forcing function,
\EQ
\ff_{\kk}^{\rm(nohel)}=
\left(\kk\times\eee\right)/\sqrt{\kk^2-(\kk\cdot\eee)^2},
\label{nohel_forcing}
\EN
is a non-helical forcing function, and $\eee$ is an arbitrary unit vector
not aligned with $\kk$; note that $|\ff_{\kk}|^2=1$.

\section*{Acknowledgements}
We are grateful to Micha{\l} Hanasz and John Scalo for useful discussions.
We acknowledge the help of an anonymous referee in improving the presentation.
This work was supported by PPARC grants {PPA/S/S/2000/02975A},
{PPA/S/S2002/03473}, {PPA/G/S/2000/00528}.
APS, AJM and AS are grateful to
Nordita for financial support and hospitality. We acknowledge the Danish
Center for Scientific Computing for granting time on the Horseshoe cluster.


\label{lastpage}

\begin{thebibliography}{99}

\bibitem[\protect\citeauthoryear{Bakunin}{2003a}]{B03a}
Bakunin, O.~G. 2003a, Uspekhi Fiz.\ Nauk, 173, 317 (Physics-Uspekhi, 46, 309)

\bibitem[\protect\citeauthoryear{Bakunin}{2003b}]{B03b}
Bakunin, O.~G. 2003b, Uspekhi Fiz.\ Nauk, 173, 757 (Physics-Uspekhi, 46, 733)

\bibitem[\protect\citeauthoryear{Berezinskii et al.}{1990}]{Berezinskiietal90}
Berezinskii, V.~S., Bulanov, S.~V., Dogiel, V.~A., Ginzburg, V.~L., \&
Ptuskin, V.~S. 1990, Astrophysics of Cosmic Rays (V.~L.~Ginzburg, ed.),
Amsterdam, North-Holland

\bibitem[\protect\citeauthoryear{Blackman \& Field}{2003}]{BF03}
Blackman, E.~G., \& Field, G.~B.\ypf{2003}{15}{L73}

\bibitem[\protect\citeauthoryear{Barge \& Sommeria}{1995}]{BS95}
Barge, P., \& Sommeria, J.\yana{1995}{295}{L1}

\bibitem[\protect\citeauthoryear{Brandenburg}{2001}]{B01}
Brandenburg, A.\yapj{2001}{550}{824}

\bibitem[\protect\citeauthoryear{Brandenburg \& Dobler}{2001}]{BD01}
Brandenburg, A., \& Dobler, W.\yana{2001}{369}{329}

\bibitem[\protect\citeauthoryear{Brandenburg \& Subramanian}{2005}]{BS05}
Brandenburg, A., \& Subramanian, K.\ 2005, Phys.\ Rep., 417, 1

\bibitem[\protect\citeauthoryear{Brandenburg et al.}{1995}]{BPS95}
Brandenburg, A., Procaccia, I., \& Segel, D.\ypp{1995}{2}{1148}

\bibitem[\protect\citeauthoryear{Brandenburg et al.}{2004}]{BKM04}
Brandenburg, A., K\"apyl\"a, P., \& Mohammed, A.\ypf{2004}{16}{1020}

\bibitem[\protect\citeauthoryear{Cesarsky}{1980}]{C80}
Cesarsky, C.~J.\yaraa{1980}{18}{289}

\bibitem[\protect\citeauthoryear{Chuvilgin \& Ptuskin}{1993}]{CP93}
Chuvilgin, L.~G., \& Ptuskin, V.~S., \yana{1993}{279}{278}

\bibitem[\protect\citeauthoryear{Drury \& V\"olk}{1981}]{DV81}
Drury, L. O'C., \& V\"olk, J. H.\yapj{1981}{248}{344}

\bibitem[\protect\citeauthoryear{Elperin et al.}{1996}]{EKR96}
Elperin, T., Kleeorin, N., \& Rogachevskii, I.\yprl{1996}{77}{5373}

\bibitem[\protect\citeauthoryear{Elperin et al.}{1997}]{EKR97}
Elperin, T., Kleeorin, N., \& Rogachevskii, I.\ypre{1997}{55}{2713}

\bibitem[\protect\citeauthoryear{Farmer \& Goldreich}{2004}]{FG04}
Farmer, A.~J., \& Goldreich, P.\yapj{2004}{604}{671}

\bibitem[\protect\citeauthoryear{Felice \& Kulsrud}{2001}]{FK01}
Felice, G.~M., \& Kulsrud, R.~M.\yapj{2001}{553}{198}

\bibitem[\protect\citeauthoryear{Gombosi et al.}{1993}]{GombosiEtal93}
Gombosi, T.~I., Jokipii, J. R., Kota, J., Lorencz, K., \& Williams, L. L.\yapj{1993}{403}{377}

\bibitem[\protect\citeauthoryear{Hanasz \& Lesch}{2003}]{HanaszLesch2003}
Hanasz, M. \& Lesch, H.\yana{2003}{412}{331}

\bibitem[\protect\citeauthoryear{Hanasz et al.}{2004}]{HanaszEtal2004}
Hanasz, M., Kowal, G., Otmianowska-Mazur, K., \& Lesch, H.\yapj{2004}{605}{L33}

\bibitem[\protect\citeauthoryear{Haugen et al.}{2004}]{Haugen04}
Haugen, N. E. L., Brandenburg, A., \& Mee, A. J.\ymn{2004}{353}{947}

\bibitem[\protect\citeauthoryear{Hillas}{2005}]{Hil05}
Hillas, A. M.\yjour{2005}{J.\ Phys.\ G: Nucl.\ Part.\ Phys.}{31}{R95}

\bibitem[\protect\citeauthoryear{Hodgson \& Brandenburg}{1998}]{HB98}
Hodgson, L. S. \& Brandenburg, A.\yana{1998}{330}{1169}

\bibitem[\protect\citeauthoryear{Johansen et al.}{2004}]{JAB04}
Johansen, A., Andersen, A. C., \& Brandenburg, A.\yana{2004}{417}{361}

\bibitem[\protect\citeauthoryear{Jun et al.}{1994}]{JCN94}
Jun, B.-I., Clarke, D. A., \& Norman, M. L.\yapj{1994}{429}{748}

\bibitem[\protect\citeauthoryear{Kang \& Jones}{1990}]{KJ90}
Kang, H., \& Jones, T. W.\yapj{1990}{353}{149}

\bibitem[\protect\citeauthoryear{K\'ota \& Jokipii}{2000}]{KJ00}
K\'ota, F., \& Jokipii, J.\ R.\yapj{2000}{531}{1067}

\bibitem[\protect\citeauthoryear{Krause}{1972}]{Krause1972}
Krause, F.\yan{1972}{294}{83}

\bibitem[\protect\citeauthoryear{Krause \& R\"adler}{1980}]{KR80}
Krause, F., \& R\"adler, K.-H.\ybook{1980}
{Mean-Field Magneto\-hydrodynamics and Dynamo Theory}
{Akademie-Verlag, Berlin; also Pergamon Press, Oxford}

\bibitem[\protect\citeauthoryear{Landau \& Lifshitz}{1987}]{LL87}
Landau, L.~D., \& Lifshitz, E.~M.\ybook{1987}
{Fluid Mechanics}
{2nd Edition, Pergamon Press, Oxford}

\bibitem[\protect\citeauthoryear{Mac Low \& Klessen}{2004}]{MLK04}
Mac Low, M.-M., \& Klessen, R.~S., 2004, Rev.\ Mod.\ Phys., 76, 125

\bibitem[\protect\citeauthoryear{Miller \& Stone}{2000}]{MillerStone00}
Miller, K.\ A. \& Stone, J.\ M.\yapj{2000}{534}{398}

\bibitem[\protect\citeauthoryear{Moss et al.}{1999}]{MSS99}
Moss, D., Shukurov, A., \& Sokoloff, D.\yana{1999}{343}{120}

\bibitem[\protect\citeauthoryear{Padoan \& Scalo}{2005}]{PadoanScalo05}
Padoan, P., \& Scalo, J.\yapjl{2005}{624}{L97}

\bibitem[\protect\citeauthoryear{Parker}{1966}]{Parker1966}
Parker, E.~N.\yapj{1966}{145}{811}

\bibitem[\protect\citeauthoryear{Parker}{1992}]{Parker1992}
Parker, E.~N.\yapj{1992}{401}{137}

\bibitem[\protect\citeauthoryear{Ptuskin}{1979}]{P79}
Ptuskin, V.~S. 1979, ApSS, 61, 359

\bibitem[\protect\citeauthoryear{Ryu et al.}{2003}]{RYU}
Ryu, D., Kim, J., Hong, S.~S., \& Jones, T.~W. \yapj{2003}{589}{338}

\bibitem[\protect\citeauthoryear{Schlickeiser \& Lerche}{1985}]{SL85}
Schlickeiser, R., \& Lerche, I. \yana{1985}{151}{151}

\bibitem[\protect\citeauthoryear{Skilling}{1975}]{Skilling}
Skilling, J.\ymn{1975}{173}{255}

\bibitem[\protect\citeauthoryear{Subramanian}{1999}]{S99}
Subramanian K., 1999, Phys.\ Rev.\ Lett., 83, 2957

\bibitem[\protect\citeauthoryear{Zeldovich et al.}{1990}]{Z90}
Zeldovich, Ya.~B., Ruzmaikin, A.~A., \& Sokoloff, D.~D.\ybook{1990}
{The Almighty Chance}
{World Scientific, Singapore}

\end{thebibliography}
\end{document}